\documentclass[epj]{svjour}

\usepackage{graphicx}
\usepackage{psfrag}

\newcommand{\eq}[1]{{(\ref{#1})}}
\newcommand{\cM}{{\cal M}}
\newcommand{\td}{{\tilde d}}
\newcommand{\tu}{{\tilde u}}
\newcommand{\hf}{{\frac{1}{2}}}

\begin{document}
\title{
Supersymmetry beyond minimal flavour violation\thanks{
To appear in the EPJC special volume ``Supersymmetry on the Eve of the LHC'',
 dedicated to the memory of Julius Wess}}
\subtitle{}
\author{S. J\"ager\inst{1}
\thanks{\emph{Email:} sebastian.jaeger@cern.ch}
}
\institute{
Theory Division, Physics Department,
CERN, CH-1211 Geneva 23, Switzerland
{}
{\normalsize\normalfont{\vskip-4cm\hfill CERN-PH-TH/2008-176}\vspace*{3.5cm}}
}
%
\date{}
\abstract{
We review the sources and phenomenology of non-minimal flavour
violation in the MSSM. We discuss in some detail the most important
theoretical and experimental constraints, as well as promising
observables to look for supersymmetric effects at the LHC and in the future.
We emphasize the sensitivity of flavour physics to the mechanism
of supersymmetry breaking and to new degrees of freedom present
at fundamental scales, such as the grand unification scale.
We include a discussion of present data that may hint at departures 
from the Standard Model. 
} 
\maketitle
%

\section{Introduction}
\label{sec:intro}
The Standard Model (SM) of particle physics has performed much better
than could have been expected just after its establishment in the
1970s as a (conceptually) unified framework of all of particle physics
based on the principle of quantum field theory with
spontaneously broken gauge invariance:
It has since then provided an economical description of thousands
of measurements in particle physics and has so far eluded
falsification. Nevertheless, there are known good reasons
to believe in new physics, specifically supersymmetry (SUSY).
Among them are:

\begin{itemize}
  \item gauge coupling unification in the Minimal Supersymmetric
      Standard Model (MSSM), unlike the SM
  \item neutrino masses (and perhaps leptogenesis) from a seesaw
     mechanism at the GUT scale
  \item stabilization of the hierarchy
     $M_W \ll M_{\{\rm seesaw, GUT, Pl\}}$
  \item dark matter candidate, once proton stability is enforced by
     $R$-parity
\end{itemize}
If SUSY stabilizes the weak scale, we will know soon:
the LHC has been designed to directly probe the TeV scale and clarify the
mechanism of electroweak symmetry breaking.
ATLAS and CMS should detect at least the lightest particle
in the Higgs sector and likely part of the remaining superparticle
spectrum directly. On the other hand, the examples given above
demonstrate that precise measurements of low-energy observables
can probe fundamental
scales, including those beyond the ``energy frontier''. This happens,
for instance, if the fundamental physics violates accidental
symmetries of the low-energy theory, i.e.\ lepton flavour as well as
lepton number in the case of neutrino oscillations in the context
of the seesaw mechanism.
Hadronic flavour transitions do occur in the SM, but they
follow a very non-generic pattern as they are constrained by the
$V-A$ structure of weak interactions and the absence of
flavour-changing neutral currents (FCNC) at tree level. Moreover, they
are suppressed by the scale of electroweak symmetry breaking,
as well as the hierarchical structure of the
Cabibbo-Kobayashi-Maskawa matrix which accounts for all flavour
violation and all observed CP violation in the Standard Model.
In consequence, there exist flavour-changing weak processes, such
as neutral meson-antimeson mixing and certain weak decays, which are
suppressed by multiple factors,
\begin{eqnarray*}
 &&  V_{ti} V_{tj}^* \times \frac{1}{16 \pi^2} \times \frac{m^2}{M_W^2} \nonumber \\[3mm]
 &&\sim 10^{-2} \times 10^{-2} \times 10^{-3} 
         \qquad \mbox{for FCNC $B_d$ decays},
\end{eqnarray*}
where $m$ denotes the mass of the decaying particle.
Comparing this to the corresponding factor $ \frac{m^2}{M^2} $
due to a new flavoured particle with
{\em generic} couplings and mass $M$, we see that even if
$M \sim 10^{(3-4)} M_W \sim 10^2$ TeV, it can give rise to ${\cal O}(1)$
corrections to $B$-physics: Flavour physics is sensitive to very
large scales. Indeed, such high sensitivity might seem too much: the
absence of clear deviations from the SM pattern would seem to imply
an unnaturally large scale of new physics (NP). Nevertheless,
the absence of new flavoured degrees of freedom is unlikely
to be an option in a theory of the weak scale, as the quadratic
divergences of the Higgs/$W$ (mass)$^2$ are caused, in part, by
fluctuations of a flavoured strongly interacting particle (the top).
This guise of a ``little hierarchy'' is often called the
new-physics flavour puzzle. Similar issues exist in the lepton
sector, if new lepton-flavoured degrees of freedom are present (as in the MSSM).
Of course, the flavour puzzle looks less severe for new physics
that entails the same loop suppression as in the Standard Model.
This is the case in the MSSM. Nevertheless,
present flavour-physics data imposes strong constraints on
the SUSY flavour structure. Finally, the data do show certain
patterns that are consistent with deviations from the SM in
``reasonable places'', i.e.\ NP-sensitive loop-dominated processes,
albeit the significance is not (yet?) very high.

The remainder of this article is organized as follows.
Section \ref{sec:fv} deals with sources of SUSY flavour violation,
the connection with supersymmetry breaking, the generic patterns
that are expected depending on how SUSY is broken (and mediated),
and constraints from internal consistency of the MSSM.
Section \ref{sec:pheno} is devoted to the most important
observables that either impose constraints on the MSSM flavour
parameters at present or are likely to show signals in the future.
We review and in some cases update bounds on ``mass-insertion''
parameters commonly used in the literature.
There exist a number of articles devoted to this issue, foremost
of all we would like to mention the original work of Gabbiani
et al.\ \cite{Gabbiani:1996hi} and the review article of Misiak,
Pokorski, and Rosiek \cite{Misiak:1997ei}, on both of which we
have drawn considerably.
We next discuss, in Section \ref{sec:susygut}, some correlation patterns
that one may expect in certain SUSY GUTs, which
illustrates the power to probe very high scales by combining
information from different indirect observables.
Finally, in Section \ref{sec:bsmhints} we mention several cases of
observables where presently
patterns of (mild) deviations from the SM are seen. All of them involve
$b \to s$ transitions, which are precisely the domain of the
LHCb experiment at CERN.

\section{SUSY flavour violation}
\label{sec:fv}
\subsection{The unbroken MSSM is minimally flavour-violating}
\label{sec:lag}

The MSSM (see \cite{Nilles:1983ge,Haber:1984rc,Martin:1997ns} 
for reviews) stabilizes the
weak scale by pairing bosons and fermions and relating their couplings
to ensure a systematic Bose-Fermi cancellation, eliminating quadratic
sensitivity to the cutoff (or scale of UV completion). In the MSSM,
each chiral fermion is accompanied by a complex scalar ``sfermion'',
each gauge boson by a Weyl ``gaugino'', and each of two higgs scalar(s) by
``higgsinos'' (see Table \ref{tab:multiplets}).
\begin{table}
\caption{Supermultiplets in the MSSM (only propagating component
fields are listed). Superpartners are denoted by a tilde.
The last column denotes how often a multiplet
appears.}
\label{tab:multiplets}       
\centerline{
\begin{tabular}{ccccccc}
\hline\noalign{\smallskip}
multiplet & \multicolumn{3}{c}{components} & SM gauge group
& $R$-parity & \# \\
& \multicolumn{3}{c}{of spin} & & & \\
& $0$ & $\hf$ & $1$ & representation &\\
\noalign{\smallskip}\hline\noalign{\smallskip}
$Q$ & $\tilde q$ & {\bf $q$} & &  $(3,2;1/6)$ & $-$ ($-$,$+$) & 3 \\
$U^c$ & $\tilde u^c$ & $u^c$ & & $(\bar 3,1;-2/3)$ & $-$ ($-$,$+$) & 3 \\
$D^c$ & $\tilde d^c$ & $d^c$ & & $(\bar 3,1;1/3)$ & $-$ ($-$,$+$) & 3 \\
$L$ & $\tilde l$ & $l$ & & $(1,2;-1/2)$ & $-$($-$,$+$) & 3 \\
$E^c$ & $\tilde e^c$ & $e^c$ & & $(1,1;1)$ & $-$($-$,$+$) & 3 \\
$H_d$ & $h_d$ & $\tilde h_d$ & & $(1,2;-1/2)$ & $+$($+$,$-$) & 1 \\
$H_u$ & $h_u$ & $\tilde h_u$ & & $(1,2;1/2)$ & $+$($+$,$-$) & 1 \\
$V_G$ & & $\tilde g$ & $G$ & $(8,1;0)$ & $+$($-$,$+$) & 1 \\
$V_W$ & & $\tilde w$ & $W$ & $(1,3;0)$ & $+$($-$,$+$) & 1 \\
$V_B$ & & $\tilde b$ & $B$ & $(1,1;0)$ & $+$($-$,$+$) & 1 \\
\noalign{\smallskip}\hline
\end{tabular}
}
\end{table}

With new flavoured degrees of freedom, one generically expects
modified flavour physics. However, the supersymmetrization itself
does not introduce any new flavour structures.
Indeed, in the limit where supersymmetry breaking is
switched off, all flavour violation resides in the
superpotential
($a \cdot b = a_1 b_2 - a_2 b_1$ is the invariant $SU(2)$ bilinear;
our notation for the couplings conforms to the SUSY Les Houches accord
conventions \cite{Allanach:2008qq} wherever it overlaps):
\begin{eqnarray} \label{eq:superpotential}
   W &=& \mu H_u \cdot H_d
      + Y^U_{ij} Q_i \cdot H_u U^c_j \nonumber \\
&&      + Y^D_{ij} H_d \cdot Q_i D^c_j
      + Y^E_{ij} H_d \cdot L_i E^c_j .
\end{eqnarray}
Two higgs doublets $H_u$, $H_d$ are required by gauge anomaly cancellation
in the presence of the higgsinos. They are also necessary for fermion mass
generation, as supersymmetry implies that each doublet has a
well-defined (sign of the) hypercharge, hence can give mass to either
$T_3=+1/2$ or to $T_3=-1/2$ SM fermions but not both. Thus in spite
of the extra doublet, there are only three Yukawa terms, as many and
as fundamental as in the SM. This relegates the origin of flavour breaking
to more fundamental scales (an important difference to
technicolour theories or extra-dimensional setups with bulk fermions).

Unlike in the SM, $B$ and $L$ are not accidental symmetries at the
renormalizable level. In writing \eq{eq:superpotential} we 
have omitted additional $B$ or $L$-violating terms such as $U^c_i D^c_j D^c_k$
or $E^c_i L_j\cdot L_k$. The former, for example, generically mediates
proton decay at unacceptably large rates. The extra terms are absent
from \eq{eq:superpotential} if $R$-parity is conserved. 
This also makes the lightest superpartner stable and restricts
superpartners to only appear in loops in low-energy processes that involve
only external SM particles.\footnote{The
behaviour of the fields under an $R$-parity
transformation (reflection of superspace coordinates) is shown in
the fourth column of Table \ref{tab:multiplets}. All ``SM'' particles
(including both scalar higgs doublets) are even, all superpartners odd.}

The Lagrangian follows from the superpotential as
\begin{eqnarray}  \label{eq:LfromW}
  {\cal L} &=& \int d^4 \theta K(\phi, \phi^*) +
       \Big\{ \int d^2 \theta\, W(\phi)
        +   \mbox{h.c.} \Big\}
\nonumber \\  && + \mbox{gauge kinetic terms} .
\end{eqnarray}
At the renormalizable level, the K\"ahler potential $K$ is fixed to
the form
\begin{eqnarray}
   K(\phi, \phi^*) &=& \sum_i \phi_i^* e^{2 g_a V_a} \phi_i .
\end{eqnarray}
The interactions among the fermions and sfermions are then given in terms
of $W$ and gauge couplings as
\begin{eqnarray}
  {\cal L}_{\rm Yukawa} &=& \frac{1}{2} \sum_{ij} \frac{\partial^2 W}{\partial
    \phi_i \partial \phi_j} \psi_i^T C \psi_j + \mbox{h.c.},
                 \label{eq:yukawapotential} \\
  {V}(\{\phi_i\}) &=& \sum_i \Bigg| \frac{\partial W}{\partial \phi_i} \Bigg|^2
    + \frac{1}{2} \sum_{aA} g_a^2 (\sum_i \phi^\dagger_i T^A_a
    \phi_i)^2 , \label{eq:scalarpotential}
\end{eqnarray}
and the fermion-sfermion-gaugino interactions are
proportional to gauge couplings.
This leaves the MSSM with one parameter {\em fewer} than
the Standard Model. (The missing parameter is the quartic Higgs
coupling, which is fixed in terms of gauge couplings.)

On the flavour side, the gauge interactions respect an $SU(3)^5$ flavour
symmetry -- separate transformations
\begin{equation} \label{eq:flavoursym}
   \phi_i \to U^\phi_{ij} \phi_j \qquad \mbox{($\phi = Q, U^c, D^c, L, E^c$)}
\end{equation}
among the three generations of any of the five types of irreducible
supermultiplets -- which
is broken only by $Y^U, Y^D, Y^E$. Hence at this stage, the
MSSM is minimally flavour violating
\cite{Hall:1990ac,Buras:2000dm,Buras:2000qz,D'Ambrosio:2002ex}. 
Indeed, once electroweak symmetry is broken by vacuum expectation values
\begin{equation}
  \langle h_u \rangle = \left( \matrix{ 0 \cr \frac{v_u}{\sqrt{2}} } \right) ,
  \qquad
  \langle h_d \rangle = \left( \matrix{ \frac{v_d}{\sqrt{2}} \cr 0 } \right)
\end{equation}
(which are real for a suitable phase convention for $h_u$, $h_d$),
the mass matrix of the down-type
sfermions in the so-called super-CKM basis\footnote{
The super-CKM basis  (of superfields) \cite{Dugan:1984qf,Hall:1985dx}
is defined by requiring diagonal Yukawa couplings in
\eq{eq:superpotential} and the CKM matrix being in its
four-parameter standard form.
} is compactly expressed in terms of $3\times 3$ blocks as\footnote{
The order of rows and columns is:
$(\tilde d_L, \tilde s_L, \tilde b_L, \tilde d_R, \tilde s_R, \tilde
b_R)$, where $\tilde d_{iL} \equiv \tilde d_i$,
$\tilde d_{iR} \equiv \tilde (d_i^c)^*$, for $\cM^2_{\tilde d}$,
with obvious analogy for up-type squarks and charged sleptons.}
\begin{eqnarray}
  \cM^{2, \rm sym}_{\tilde f} &\equiv&
\left( \matrix{
   \cM^{2, \rm sym}_{\tilde f LL} &
   \cM^{2, \rm sym}_{\tilde f LR}
\cr \cr
  \big( \cM^{2, \rm sym}_{\tilde f LR} \big)^\dagger &
	\cM^{2, \rm sym}_{\tilde f RR} } 
\right) \nonumber \\[2mm]
   &\equiv&
\left( \matrix{
  m^2_f + D_{fLL} &
  - \mu\, m_f\, x_f
\cr \cr
  - \mu^*\, m_f\, x_f &
	m^2_f + D_{fRR} } \label{eq:m2sym}
\right) .
\end{eqnarray}
Here the label $f=u,d,e$ denotes up-type squarks, down-type squarks, 
and charged sleptons, respectively,
$m_u$, $m_d$, and $m_e$ are the corresponding $3 \times 3$ {\em fermion} mass
matrices, and $x_u=\cot\beta\equiv v_d/v_u$ and $x_d=x_e=\tan\beta=v_u/v_d$.
The $D$-term contributions $D_{fLL(RR)}$ are proportional to
the $3\times3$ unit matrix, i.e., flavour-blind:
\begin{eqnarray}
  D_{fLL,RR} &=& \cos (2\beta)\, m_Z^2\, (T_{3f} - Q_f \sin^2 \theta_W )\, {\bf
    1} \nonumber \\
  &\equiv& d_{fLL,RR} \cos (2\beta)\, m_Z^2\, {\bf 1} .
\end{eqnarray}
The values $d_{fLL,RR}$ are reported in Table \ref{tab:dterms}.
\begin{table}
\caption{$D$-term coefficients $d_{fLL,RR}$}
\label{tab:dterms}       
\centerline{
\begin{tabular}{lccc}
\hline\noalign{\smallskip}
 & $u$ & $d$ & $e$  \\
\noalign{\smallskip}\hline\noalign{\smallskip}
$LL$ & $\frac{1}{2} - \frac{2}{3} \sin^2 \theta_W$ &
  $-\frac{1}{2} + \frac{1}{3} \sin^2 \theta_W$  &
  $-\frac{1}{2} + \sin^2 \theta_W$ \\
$RR$ & $\frac{2}{3} \sin^2 \theta_W$ &
  $-\frac{1}{3} \sin^2 \theta_W$ &
  $-\sin^2 \theta_W$ \\
\noalign{\smallskip}\hline
\end{tabular}
}
\end{table}
Clearly, the sfermion mass matrices are diagonal in the super-CKM basis.
In other words, the flavour sfermion states coincide with sfermion
mass eigenstates. We note that the winos, binos,
and higgsinos shown in the vertices are not mass eigenstates, but
rather mix to form charginos and neutralinos. However, this does not
affect the flavour structure of the vertices.
The fact that sfermion mass and flavour states coincide
implies that flavour violation is
exclusively due to (s)fermion-charge-changing interactions. Those that
involve fermions take
the form of charged-wino-quark-squark, charged-higgsino-quark-squark,
and charged-Higgs-quark-quark couplings, besides the
$W u d$ and $W l \nu$ couplings in the Standard Model (Fig. \ref{fig:fvsckm}).
\begin{figure}
\vskip-2cm
\hspace*{-3.5cm}
\includegraphics[width=15cm,angle=0]{fv_fbasis.eps}
\caption{Flavour-changing vertices involving fermions
in the super-CKM basis.
\label{fig:fvsckm}       
}
\end{figure}
We note that all flavour violation at vertices is governed in the
obvious way by the CKM matrix. In the figure, we have switched to
four-component fermion notation. Furthermore we have defined
\begin{equation}
 \tilde f_{R} \equiv (\tilde f^c)^* .
\end{equation}
The left-handed quark flavour states are related to the weak doublets
(``interaction basis'') via
\begin{equation}
   \tilde q_{Li} = (V_{ki}^* \tilde u_{Lk}, \tilde d_{Li}),
\end{equation}
where, as is common, we have selected out of the infinite set of
interaction bases one where the down-type Yukawa couplings are
diagonal.\footnote{A model of flavour would single out a(t least one)
different basis (in general) where the Yukawa matrices would look
``simple''.}
The sneutrino flavour states are defined as $SU(2)$ and SUSY partners of the
charged-lepton mass eigenstates,
\begin{equation}
   \tilde l_{Li} = (\tilde \nu_{Li}, \tilde e_{Li}) .
\end{equation}
This minimal set of flavour-violating vertices immediately
leads to a very non-generic pattern of (quark-) flavour transitions.
In particular, $b\to s$, $b\to d$, and $s \to d$ transitions
are corrrelated, as they are in the SM \cite{Buras:2000qz}.
The nonminimal flavour violation that is the title of this
article is then exclusively due to supersymmetry breaking.
Strictly speaking, our title is an oxymoron: it should really read
``Flavour violation beyond the SUSY limit''!
This makes clear that the search for departures from minimal flavour violation
provides a line of approach in unraveling the
SUSY breaking mechanism.

\subsubsection{$\tan\beta$ parameter}
The parameter $\tan\beta=v_u/v_d$ introduced below \eq{eq:m2sym} plays an
important role in Higgs physics, as it affects both the mass spectrum
and the trilinear couplings. The tree-level
relation $m_b = y_b v_d/\sqrt{2}$ implies that $y_b \sim y_t \sim 1$
becomes possible for $\tan\beta = {\cal O}(50)$ (as in certain grand
unified models that entail bottom-top Yukawa unification).
The relevance to flavour physics derives from
the fact that this can lead to double enhancement factors in front of
loops of the form
\cite{Hall:1993gn,Hempfling:1993kv,Carena:1994bv,Blazek:1995nv,Hamzaoui:1998nu,Choudhury:1998ze,Babu:1999hn},
$$
      \tan\beta\, y_b \propto \tan^2\beta\, m_b , 
$$
which can compensate the loop suppression in Higgs-mediated
contributions to FCNC observables
which are usually small for small to moderate ($<30$) values of
$\tan\beta$ but can give rise to a distinctive pattern at larger
values even for minimal flavour violation.
We will not discuss these effects; for a recent
review see \cite{Isidori:2007ed}. Most of the constraints discussed
below still apply in that case, but there may be stronger ones.

\subsection{Origin of (new) flavour violation: supersymmetry breaking}
The superpotential \eq{eq:superpotential} does not break supersymmetry
spontaneously at tree level. Because of supersymmetric
nonrenormalization theorems \cite{Zumino:1974bg,Grisaru:1979wc,Seiberg:1993vc},
this remains true to all orders in perturbation theory.
Neither is electroweak symmetry broken, at any order.

Observations exclude the presence of mass-degenerate superpartners
for many of the SM particles, which tells us that
supersymmetry is broken.
The standard picture is that supersymmetry
breaking occurs in a hidden sector of SM gauge singlets,
via the condensation of an auxiliary ($F$ or $D$) component
of one or more superfields $X$.
Gauge symmetry then requires any superpotential couplings between the
visible and hidden sectors to be nonrenormalizable.\footnote{The 
one exception is a possible coupling $H_u \cdot H_d X$,
without imposing further global symmetries.}
In many cases of interest, all low-energy effects of supersymmetry breaking
can be represented by such effective nonrenormalizable superpotential,
gauge-kinetic, and K\"ahler terms, as in
\begin{equation} \label{eq:higherdimW}
   W_{\rm break} = 
       {\cal A}^U_{ij} \frac{\langle X \rangle}{M} U^C_i H_u \cdot Q_j ,
\end{equation}
\begin{equation} \label{eq:higherdimf}
   f_{\rm break} =
       {\cal M}_a \frac{\langle X \rangle}{M} W_a^A W_a^A ,
\end{equation}
and
\begin{equation} \label{eq:higherdimK}
   K_{\rm break} =  {\cal K}^Q_{ij} \frac{\langle X X^\dagger
     \rangle}{M^2} Q_i^\dagger e^{2 g_a V_a} Q_j .
\end{equation}
Here ${\cal A}^U_{ij}$, ${\cal M}_a$, and ${\cal K}^Q_{ij}$ are
dimensionless coefficients. $\langle X \rangle = \theta^2 F_X$
is the vacuum expectation value of a hidden-sector superfield, and
the SUSY-breaking terms in the Lagrangian are found by replacing
$K \to K + K_{\rm break}$ and
$W \to W + W_{\rm break} + f_{\rm break}$ in \eq{eq:LfromW}.
This can be illustrated as follows. The MSSM, by assumption, does not
have any direct renormalizable couplings to the hidden sector. Assume
then that the lightest ``messenger'', i.e., degree of freedom that
couples both to the field $X$ and to the MSSM fields, has mass $M$.
Below its mass scale, it can be
integrated out of the theory, giving rise to operators
as in \eq{eq:higherdimW}--\eq{eq:higherdimK}.
This is what happens, for example, in models of gauge mediation (see below).

The term $W_{\rm break}$ from above gives rise to an extra contribution
$$
   {\Delta \cal L}_{A} =
       T^U_{ij}\, \tilde q_i \cdot h_u \tilde u_j^c + \mbox{h.c.},
$$
\begin{equation} \label{eq:effA}
   T^U_{ij} = \frac{F_X}{M} {\cal A}^U_{ij}
\end{equation}
to the trilinear scalar coupling (traditionally called ``$A$-term'').
On the other hand, \eq{eq:higherdimf} generates gaugino masses,
while \eq{eq:higherdimK} gives rise to extra contributions
$$
   {\Delta \cal L}_{m^2} = m^2_{\tilde qij} \, \tilde q_i^\dagger \tilde q_j ,
$$
\begin{equation} \label{eq:effm}
   m^2_{\tilde qij} = \frac {F_X^2}{M^2} {\cal K}^Q_{ij}
\end{equation}
to the masses of the sfermions (relative to those of their fermionic partners).

The crucial point for flavour
is the following: Barring any supplement of the
MSSM by a theory of flavour, naturalness dictates that
the flavour structures of $m^2_{\tilde qij}$ and $T^U$ should be assumed generic
numbers ${\cal O}(1)$ and, in particular, independent of $Y^U$.
Hence the SUSY-breaking, renormalizable (masses and) interactions
among the sfermions provide a generically nonminimal source of flavour
violation.
What can one say about the mass scale $M$? First, \eq{eq:effm}
shows that the effective SUSY-breaking mass scale is set by
$$
   m_{\rm SUSY} = F_X/M
$$
which should be of ${\cal O}(\rm TeV)$ to preserve SUSY as a solution
to the hierarchy problem, thus $M$ can in principle
range from not far beyond the SUSY scale.
On the other hand, any global flavour symmetries which would
forbid or restrict the nonrenormalizable terms of the
gauge sector are expected to be broken (nonminimally) by gravitational
physics, such that $M < M_{Pl}$.

Generalizing \eq{eq:higherdimW} and \eq{eq:higherdimK}
and including operators as in \eq{eq:higherdimf} that give rise to
gaugino masses induces the following set of dimensionful SUSY-breaking terms:
\begin{eqnarray}
{\cal L}_{\rm soft} &=&
    - m^2_{\tilde q ij} \tilde q^\dagger_i \tilde q_j
    - m^2_{\tilde u ij} \tilde u^{c\dagger} \tilde u^c
    - m^2_{\tilde d ij} \tilde d^{c\dagger} \tilde d^c
\\
 && - m^2_{\tilde l ij} \tilde l^{\dagger} \tilde l
    - m^2_{\tilde e ij} \tilde e^{c\dagger} \tilde e^c
    - m^2_{h_U} h_u^\dagger h_u
    - m^2_{h_d} h_d^\dagger h_d
\nonumber  \\
 && - \Big[ m_1 \tilde b \tilde b + m_2 \tilde w^A \tilde w^A
    + m_3 \tilde g^A \tilde g^A + B_\mu h_u \cdot h_d
\nonumber \\
 && + T^U_{ij} \tilde q_i \cdot h_u \tilde u_j^c
 + T^D_{ij} h_d \cdot \tilde q_i \tilde d_j^c
 + T^E_{ij} h_d \cdot \tilde l_i \tilde e_j^c  + \mbox{h.c.} \Big] .
\nonumber
\end{eqnarray}
${\cal L}_{\rm soft}$ does not introduce any quadratic divergences.
In particular, the soft-breaking terms themselves are only
logarithmically sensitive to heavy scales such as the seesaw scale.

From a purely phenomenological point of view, the supersymmetry
breaking can simply be introduced explicitly according to the structure
of ${\cal L}_{\rm soft}$. In this case, the hierarchy is stabilized but
not explained: all dimensionful terms in ${\cal L}_{\rm soft}$
have to simply happen to be of order the weak scale. Indeed, often
the MSSM is defined as the supersymmetrized SM with explicit soft breaking.
There are, in fact, additional dimensionful couplings that one could
write, such as
\begin{equation}    \label{eq:nonhomom}
    c_{ij}^u (h_d^\dagger \tilde q_i) \tilde u_j^c
    + c_{ij}^d (h_u^\dagger \tilde q_i) \tilde d_j^c 
    + c_{ij}^e (h_u^\dagger \tilde l_i) \tilde e_j^c .
\end{equation}
In the case of the MSSM, these terms do not reintroduce quadratic
divergences either and are not generated by
radiative corrections from ${\cal L}_{\rm soft}$. Hence, it is
consistent to set them to zero. In an expansion in $1/M$, where $M$
is the ``messenger'' scale, they are generated at higher orders.

Finally, let us remark that the soft masses $m^2_{h_u}$,
$m^2_{h_d}$, and $B_\mu$ provide (for suitable values) for
electroweak symmetry breaking, while the hierarchy $M_W \sim
M_{\rm SUSY} \ll M_X$, where $M_X$ denotes one of the
large scales $M_{\rm seesaw}$, $M_{\rm GUT}$, $M_{\rm Pl}$, is
stabilized by the softly broken supersymmetry.
More fundamentally, the vacuum expectation values $\langle X \rangle$
may be due to dynamical supersymmetry breaking in the hidden sector,
which can naturally
generate the large hierarchy $M_{\rm SUSY} \ll M_X$ by dimensional
transmutation~\cite{Witten:1981nf}.

\subsection{Patterns of SUSY breaking and their flavour}
This section describes the expected flavour structure and mass
spectrum in the two most popular mediation schemes. The first is
gravity mediation, where the fact that SUSY is broken in the hidden
sector is communicated to the MSSM particles by their gravitational
interactions, which are always present
\cite{Nilles:1982ik,Chamseddine:1982jx,Barbieri:1982eh,Cremmer:1982vy,Ibanez:1982ee,Nilles:1982dy,Hall:1983iz,Ohta:1982wn,Ellis:1982wr,AlvarezGaume:1983gj}.
 These effects include
``anomaly-med\-iat\-ed'' contributions as a subset.
The second scheme is gauge mediation, where there is an additional
``messenger'' sector containing (supersymmetrically) heavy particles
charged under the SM gauge group and with direct couplings to
the hidden sector SUSY-breaking field(s) $X$. Here the flavour
structure is very non-generic, being flavour-blind at the
messenger scale.
\subsubsection{Gravity mediation}
Coupling a supersymmetric theory to Einstein gravity implies
invariance under local supersymmetry transformation and the presence
of a spin-3/2 gravitino field, as well as certain auxiliary fields for
the gravitational supermultiplet. Apart from that, the theory is
specified by a K\"ahler potential $K$, a superpotential $W$,  plus a
gauge kinetic function $f$, as in the nongravitational case.
Since the SUSY-mediating effects are themselves $1/M_{\rm Pl}$-suppressed,
nonrenormalizable terms have to be kept in all three functions.
Upon integrating out the supergravity auxiliary fields, couplings
beetween the hidden- and visible-sector fields are generated.
If $K$, $W$, and $f$ are generic functions of $\phi_i/M_{Pl}$, subject
only to constraints from the gauge symmetry, substituting expectation
values for the SUSY-break\-ing vevs $\langle X \rangle$ leads to
non-universal SUSY-break\-ing scalar masses. For instance,
in the case of a single hidden-sector $F$-term expectation value
$\langle X \rangle = \langle F_X \rangle \theta^2$,
there will be a contribution
\begin{eqnarray}
  \Delta V_{\rm soft} = \phi_i \phi_j^* \frac{\partial^2 k}{\partial
    \phi_i \phi_j^*} \Big|_{\phi=0},
\end{eqnarray}
where
$$
k = \frac{\partial^2 K}{\partial X \partial X^*} \Big|_{X=0} .
$$
This contribution to scalar soft masses arises from quartic terms in
the K\"ahler potential, which are $1/M_{\rm Pl}^2$-suppressed, hence
receive ${\cal O}(1)$ (relative) contributions from Planck-scale physics. There
is no a-priori reason why these should have any particular structure.
Likewise, the trilinear terms in ${\cal L}_{\rm soft}$ can have a
generic flavour structure. On the other hand, supergravity models do
not appear to lead to sizable values for the terms in \eq{eq:nonhomom}
(which arise at higher orders in $1/M_{\rm Pl}$).
In fact, although the mediation of SUSY breaking in supergravity
might be viewed as due to ``light'' particles (the gravitational
supermultiplet), all effects of supersymmetry breaking in the hidden
sector on the MSSM sector can be accounted for by higher-dimensional
operators of the form
\eq{eq:higherdimW},\eq{eq:higherdimf},\eq{eq:higherdimK},
with $M \to M_{\rm Pl}$. (In general, the effective renormalizable
visible-sector superpotential couplings will also differ from their
counterparts in $W$.)

In summary, the generic pattern for gravity mediation is the softly
broken MSSM in its full generality. More details and discussions
of specific models can be found e.g.\ in the reviews
\cite{Brignole:1997dp,Chung:2003fi}.

\subsubsection{mSUGRA}
Specific assumptions on the K\"ahler function $K$ and the breakdown
of supersymmetry in the hidden sector (such as dilaton domination
in string-theory compactifications) lead to non-generic forms.
In the popular mSUGRA (or CMSSM) scenario, it is assumed
that all running soft scalar masses unify at a certain scale
(usually, $M_{\rm GUT}$),
\begin{eqnarray} \label{eq:msugram0}
   m^2_{\tilde q ij} = m^2_{\tilde u ij} = m^2_{\tilde d ij} =
   m^2_{\tilde l ij} = m^2_{\tilde e ij}
    &=& m_0^2 \delta_{ij} , \nonumber \\
   m^2_{h_u} = m^2_{h_d} &=& m_0^2 .
\end{eqnarray}
Moreover, the gaugino masses unify,
\begin{equation} \label{eq:msugram12}
   m_{\tilde b} = m_{\tilde w} = m_{\tilde g} = m_{1/2} ,
\end{equation}
while the trilinear couplings satisfy
\begin{equation} \label{eq:msugraA}
   T^f_{ij} = a_0 Y^f_{ij} ; \qquad f=u,d,e .
\end{equation}
This leaves a predictive scenario with 4 parameters $m_0$, $a_0$,
$m_{1/2}$, and $B_\mu$ (moreover there is a constraint on $\mu$
from the observed weak scale). We emphasize that the CMSSM should be
considered as a limit which may receive important corrections.
While in certain scenarios one may hope that the CMSSM captures some
of the most important ``flavourless'' aspects of sparticle
phenomenology, such as spectra, production rates, etc., it is less
clear whether it is a good approximation for flavour physics,
as there usually are large -- although perhaps not ${\cal O}(1)$ --
corrections to its minimal
flavour structure from subdominant contributions to $K$
in specific models.

\subsubsection{Anomaly mediation}
Supergravity involves a universal class of contributions to the
soft-supersymmetry breaking terms whose form is independent of the 
details of the hidden-sector dependence of $K$, $W$
and $f$, related to anomalous breaking of scale invariance.
This results in contributions to gaugino masses, scalar masses, and
trilinear couplings which are fixed in terms of the RGE $\beta$ functions and
anomalous dimensions $\gamma$ up to one parameter
$m_{3/2}$ of order the gravitino mass \cite{Randall:1998uk,Giudice:1998xp},
\begin{eqnarray}  \label{eq:amsbm}
  m_{\tilde g_a} &=& - \frac{\beta_a}{2 g_a^2} m_{3/2} , \\
  T^f_{ij} &=& - \mu \frac{{\rm d} Y^f_{ij}}{{\rm d} \mu} m_{3/2},
   \qquad f=u,d,c  \label{eq:amsbA} \\
  m^2_{\tilde mij} &=& \frac{1}{2} \left(     \label{eq:amsbm2}
                    \beta_a \frac{\partial \gamma^m_{ij}}{\partial g_a}
                    + \sum_{fkl} \mu \frac{{\rm d} Y^f_{kl}}{{\rm d} \mu}
                      \frac{\partial \gamma^m_{ij}}{\partial Y^f_{kl}}  
                   \right) |m_{3/2}|^2 \nonumber , \\
&& \qquad \qquad \qquad \qquad m=q,u,d,l,e
\end{eqnarray}
(with analogous contributions to the soft masses and the $B_\mu$
parameter for the Higgs scalars).
Their form is RG invariant, and there
exist schemes where their form holds true to all orders.
It is evident from \eq{eq:amsbm}, \eq{eq:amsbA}, \eq{eq:amsbm2}
that in a perturbative theory such as
 the MSSM the contributions to $m_{\tilde g_a}$
and $T^f$ arise at one loop, while those to $m^2_{\tilde mij}$ arise at two
loops. I.e. all masses are of the same (loop) order and are
loop-suppressed relative to the gravitino mass.

From a flavour-perspective, these soft terms are highly non-generic.
In particular, for sfermions of the first two generations the
contributions from Yukawa terms are small. This implies that the
contributions to their trilinear scalar couplings $T$ are negligible and
those to their soft masses are degenerate.

\subsubsection{Gauge mediation}
In gauge mediation models
\cite{Dine:1993yw,Dine:1994vc,Dine:1995ag,Giudice:1998bp},
additional ``messenger'' particles
with SM gauge interactions and further couplings to the fields
that break supersymmetry are present. Typically, they receive
large masses $M$ from their couplings to the supersymmetry-breaking
field(s), which acquire vacuum expectation values of the form
$$
     \langle X \rangle = S_X + \theta^2 F_X .
$$
Below the scale $M$, the messengers
can be integrated out. If all relevant couplings are perturbative,
this gives rise to terms of
the form \eq{eq:higherdimW}, \eq{eq:higherdimf}, \eq{eq:higherdimK}
in a calculable manner.\footnote{A more general but in general
non-calculable parameterization which
applies to a class of strongly-coupled gauged mediation models has
been recently given in \cite{Meade:2008wd} }
The structure of the soft breaking terms is \cite{Dine:1995ag}
\begin{eqnarray}
  m_a &=& \frac{g_a^2}{16 \pi^2} \frac{F_X}{S_X}, \\
  m^2_{\tilde mij} &=& 2 \frac{F_X^2}{S_X^2}
             \Big( C_3 \left(\frac{\alpha_3}{4 \pi} \right)^2
                 + C_2 \left(\frac{\alpha_2}{4 \pi} \right)^2
                 +  y_m^2 \left( \frac{\alpha_1}{4 \pi} \right)^2
             \Big) \delta_{ij} , \nonumber \\
\end{eqnarray}
where $C_3=4/3$ for color triplets (zero for singlets) and $C_2=3/4$
for weak doublets (zero for singlets). Trilinear couplings $T^f_{ij}$
arise at two loops, hence are suppressed.
The flavour structure of gauge-mediated soft terms is completely
universal, i.e. sfermions of identical SM gauge charg\-es are degenerate
to leading order. This fact strongly suppresses flavour-changing
neutral currents. In particular, there are no contributions to any
FCNC process from one-loop squark-gluino or squark-neutralino diagrams,
as the flavour sfermions are mass eigenstates and
the only contributions to FCNC still arise from the vertices
shown in Fig.~\ref{fig:fvsckm}.
Radiative corrections modify this simple pattern. The dominant
(logarithmic) effects are accounted for by RG-evolving the
soft terms from the messenger scale down to the mass scale
$F_X/S_X$ of the superpartners.
We note that even in the case of gauge mediation, gravity-mediated
contributions will be present, but their relative importance will be
of
$$
    M/M_{\rm Pl} \sim S_X/M_{\rm Pl} ,
$$
such that gravity-mediated contributions are strongly suppressed for
a low messenger scale.

\subsubsection{Radiative corrections to the soft terms}
RGE effects in the MSSM modify both the spectrum and the flavour
structure of the soft terms. As far as the flavour is concerned,
no ``dangerous'' flavour structures are generated if they were
not present at the messenger scale. In particular, for
flavour-blind initial conditions as in gauge mediation or in msugra,
the RG-evolved soft terms still have a minimally flavour-violat\-ing
structure as defined in \cite{D'Ambrosio:2002ex}, and are closely
aligned with the fermion mass matrices (with near degeneracy
between the first two generations).

\subsubsection{Minimal flavour violation}
The principle of minimal flavour violation declares that the
flavour symmetry \eq{eq:flavoursym} of the gauge interactions
is only broken by the Yukawa couplings. That is, all
soft terms are functions of the Yukawa matrices such that
the Lagrangian is invariant when the Yukawas are treated
as spurions of the symmetry \cite{D'Ambrosio:2002ex}.
This allows then to show, for instance, that
the structure of minimal flavour violation is preserved
under (non-gravitational) radiative corrections,
and leads to strong correlations between different
flavour-changing phenomena \cite{Buras:2000dm,D'Ambrosio:2002ex}.

Such a setup might occur due to flavour dynamics
promoting the Yukawa matrices to dynamical fields which
are frozen to their vacuum expectation values (we know
of no concrete realization). In the MSSM it is realized for
the case of pure gauge mediation, which provides
universal soft masses and negligible trilinear couplings
at the messenger scale.

In summary, general gravity mediation and gauge mediation form
two extreme cases of complexity of flavour-violating soft terms,
which range from completely generic in (general)
gravity mediation
to minimally flavour-violating in gauge mediation (with a messenger
scale $M \ll M_{\rm Pl}$).

\subsection{General parameterizations}
A general parameterization of the soft-breaking terms that is
immediately useful in low-energy flavour phenomenology
is given by the set of three sfermion mass matrices, including
the effects from electroweak symmetry breaking, in the super-CKM
basis.
Together with the gaugino mass parameters and the dimension-two
soft terms in the Higgs sector (two of which
can be traded for the weak scale $v$ and the parameter $\tan\beta$)
they comprise the full information contained in ${\cal L}_{\rm soft}$.
Moreover, this parameterization is free from redundancies, i.e.,
all parameters are physical.\footnote{
after removing a freedom in phases of fields by choosing $m_{\tilde g}>0$,
$B_\mu>0$ real.} Each of the three mass matrices is the sum
of a supersymmetric and a soft-breaking part,
\begin{equation} \label{eq:m2decomp}
  {\cal M}^2_{\tilde f} = {\cal M}^{2, \rm sym}_{\tilde f}
                       + {\cal M}^{2, \rm break}_{\tilde f} ,
\end{equation}
where ${\cal M}^{2, \rm sym}_{\tilde f}$ has been given in \eq{eq:m2sym} and
\begin{eqnarray} \label{eq:m2breaku}
  \cM^{2, \rm break}_{\tilde u} &=&
 \left( \matrix{
    V_{\rm CKM}\, \hat m^2_{\tilde Q} V^\dagger_{\rm CKM} &
              \frac{v}{\sqrt{2}} \sin\beta\, \hat T_U^\dagger \cr
    \frac{v}{\sqrt{2}} \sin\beta\, \hat T_U & \hat m^2_{\tilde u} }
 \right) \nonumber \\
    &\equiv&
\left( \matrix{
   \cM^{2, \rm break}_{\tilde u LL} &
   \cM^{2, \rm break}_{\tilde u LR}
\cr \cr
  \big( \cM^{2, \rm break}_{\tilde u LR} \big)^\dagger &
	\cM^{2, \rm break}_{\tilde u RR} } 
\right) ,
\end{eqnarray} 
\begin{eqnarray} \label{eq:m2breakd}
  \cM^{2, \rm break}_{\tilde d} &=&
 \left( \matrix{
    \hat m^2_{\tilde Q} &
              \frac{v}{\sqrt{2}} \cos\beta\, \hat T_D^\dagger \cr
    \frac{v}{\sqrt{2}} \cos\beta\, \hat T_D & \hat m^2_{\tilde d} }
 \right) \nonumber \\
    &\equiv&
\left( \matrix{
   \cM^{2, \rm break}_{\tilde d LL} &
   \cM^{2, \rm break}_{\tilde d LR}
\cr \cr
  \big( \cM^{2, \rm break}_{\tilde d LR} \big)^\dagger &
	\cM^{2, \rm break}_{\tilde d RR} } 
\right) ,
\end{eqnarray}
\begin{eqnarray} \label{eq:m2breake}
  \cM^{2, \rm break}_{\tilde e} &=&
 \left( \matrix{
    \hat m^2_{\tilde L} &
              \frac{v}{\sqrt{2}} \cos\beta\, \hat T_E^\dagger \cr
    \frac{v}{\sqrt{2}} \cos\beta\, \hat T_E & \hat m^2_{\tilde e} }
 \right) \nonumber \\
    &\equiv&
\left( \matrix{
   \cM^{2, \rm break}_{\tilde e LL} &
   \cM^{2, \rm break}_{\tilde e LR}
\cr \cr
  \big( \cM^{2, \rm break}_{\tilde e LR} \big)^\dagger &
	\cM^{2, \rm break}_{\tilde e RR} } ,
\right) .
\end{eqnarray}
The mass matrix for sneutrinos $\cM^2_{\tilde \nu}$ is identical to
$\cM^{2, \rm break}_{\tilde e LL} = m^2_{\tilde L}$ in
the flavour basis (up to flavour-conserving $D$-terms, and
neglecting infinitesimal supersymmetric
contributions from neutrino masses). The hatted mass matrices
are identified with the quadratic terms in ${\cal L}_{\rm soft}$
in the super-CKM basis as
$ \hat m^2_{\tilde Q} = m^2_{\tilde q}$,
$ \hat m^2_{\tilde u} = (m^2_{\tilde u})^T$,
$ \hat m^2_{\tilde d} = (m^2_{\tilde d})^T$,
$ \hat m^2_{\tilde L} = m^2_{\tilde l}$,
$ \hat m^2_{\tilde e} = (m^2_{\tilde e})^T$.

Several remarks are in order.
\begin{enumerate}
\item 
The $LR$ masses are proportional to the electroweak scale.
Hence they
are suppressed by $v/M_{\rm SUSY}$ in the limit of a large
SUSY-breaking scale.
\item
We recall that the neutral fermion-sfermion-gaugino and
fermion-sfermion-higgsino
couplings are flavour-diagonal in the super-CKM basis, while the charged
couplings are governed by the CKM matrix.
However, unless assumptions about the mechanism of SUSY breaking are made,
the matrices entering \eq{eq:m2breaku}--\eq{eq:m2breake}
are in general nondiagonal, subject only to certain hermiticity conditions.
\item
The left-left sectors of the up- and down-type squark mass matrices are not
independent. In particular, at least one of them will violate flavour
as soon as either deviates from a multiple of the unit matrix.
Such deviations are already induced by leading-logarithmic corrections
proportional to $y_t^2$, which induce corrections that are aligned with
up-type quarks. Hence even in gauge mediation models, there will be
FCNC induced by gluino-squark loops \cite{D'Ambrosio:2002ex}.
\end{enumerate}

The nonminimal flavour violation is conveniently parameterized in terms
of the off-diagonal mass matrix elements,
or equivalently the ``$\delta$'' parameters,
\begin{equation}
\label{eq:def_delta}
\big( \delta^{\tilde f}_{ij} \big)_{AB} \equiv
\frac{ \Big[ \cM^{2,\rm break}_{\tilde fAB} \Big]_{ij} }{M^2} ,
\end{equation}
where $A,B=L,R$ and $M$ is a mass scale of order the sfermion mass eigenvalues.
(A popular flavour-dependent choice is to use
$M^2 = \sqrt{[({\cal M}^2_f)_{XX}]_{ii} [({\cal M}^2)_{YY}]_{jj}}$
in $(\delta^{\tilde f}_{ij})_{AB}$.)
When the $\delta$ parameters are small, it is possible to treat them
as perturbations~\cite{Hall:1985dx}. This is in most cases
justified by the constraints on them calculated in such an expansion;
however, caution must be taken
to expand to sufficiently high orders to account for all leading effects.

From the large number of flavour-violating parameters, it is evident
that the MSSM generally entails deviations from the SM predictions for
flavour-violating processes. This opens the possibility to either
observe supersymmetry indirectly or to constrain its parameters
(flavour-violating as well as flavour-conserving ones). One virtue of
flavour-violating processes is the large number of observables and the
availability of theoretical tools for rather precise predictions
for many of them. On the other hand, for generic $\delta \sim 1$ and
TeV-scale sparticle masses (such as in generic gravity-mediated
setups) experimental bounds on FCNC processed such
as $B \to X_s \gamma$ are violated (the ``SUSY flavour problem'').
However, as elaborated above, the structure of ${\cal L}_{\rm soft}$ is
intimately tied to the unknown mechanism of SUSY breaking. For instance,
gauge-mediation models
have little trouble in satisfying the low-energy constraints from
flavour physics because the SUSY breaking is tranferred by
flavour-blind gauge interactions at relatively low scales.

\subsection{Constraints from vacuum stability} \label{sec:ccbufb}
Before discussing the consequences of non-minimal flavour violation
for phenomenology (and vice versa), we note that,
independently of experimental searches for FCNC or other low-energy
SUSY-induced effects, there exists a class of ``theoretical''
constraints on the soft terms following from the requirement of
the correct electroweak symmetry breaking pattern,
which also constrain flavour-off-diagonal mass matrix elements.
The reason is that, with many scalar fields in the MSSM, there are
flat directions in field space (where the supersymmetric contribution to
the scalar potential vanishes). Unless the trilinear couplings satisfy
certain constraints, the softly broken scalar potential has charge-
or colour-breaking (CCB) minima, or is unbounded from below (UFB) in certain
directions. The authors of \cite{Casas:1996de} obtained
CCB and UFB bounds on the $(\delta^{\tilde f}_{ij})_{LR}$ , which respectively
read:
\begin{eqnarray}  \label{eq:ccbfirst}
  (\delta^{\tilde u}_{ij})_{LR} &<&
     m_{u_k} \frac{\sqrt{m^2_{\tilde u_{Li}}+m^2_{\tilde u_{Rj}}+m^2_{h_u}}}{M^2} , \\
  (\delta^{\tilde d}_{ij})_{LR} &<&
     m_{d_k} \frac{\sqrt{m^2_{\tilde d_{Li}}+m^2_{\tilde d_{Rj}}+m^2_{h_d}}}{M^2} , \\
  (\delta^{\tilde e}_{ij})_{LR} &<&
     m_{e_k} \frac{\sqrt{m^2_{\tilde e_{Li}}+m^2_{\tilde e_{Rj}}+m^2_{h_d}}}{M^2} ,
\end{eqnarray}
and
\begin{eqnarray}
  (\delta^{\tilde u}_{ij})_{LR} &<&
     m_{u_k} \frac{\sqrt{m^2_{\tilde u_{Li}}+m^2_{\tilde u_{Rj}}
                        +m^2_{\tilde e_{Li}}+m^2_{\tilde e_{Rj}}}}{M^2} , \\
  (\delta^{\tilde d}_{ij})_{LR} &<&
     m_{d_k} \frac{\sqrt{m^2_{\tilde d_{Li}}+m^2_{\tilde d_{Rj}}
                   + \frac{1}{2}(m^2_{\tilde e_{Li}}+m^2_{\tilde e_{Rj}})}}{M^2} , \\
  (\delta^{\tilde e}_{ij})_{LR} &<&  \label{eq:ufblast}
     m_{e_k} \frac{\sqrt{3 (m^2_{\tilde e_{Li}}+m^2_{\tilde e_{Rj}})}}{M^2} ,
\end{eqnarray}
where always $k=\max(i,j)$.)
Not only are these bounds in several cases stronger than those from experimental
data, but an important virtue ist that
they do not become weaker as the SUSY scale is raised; this is different
from the FCNC bounds reviewed below.
(The equivalent bounds on the trilinear $T$-parameters
themselves scale like $y M$, where $y$ is the larger of the Yukawa
coulings involved.)
\begin{table}
\caption{CCB and UFB constraints on flavour-violating
parameters $(\delta^{\tilde f}_{ij})_{LR}$  \cite{Casas:1996de}}
\label{tab:ccbufb}       
\centerline{
\begin{tabular}{lccc}
\hline\noalign{\smallskip}
 & $u$ & $d$ & $e$  \\
\noalign{\smallskip}\hline\noalign{\smallskip}
$12$ & $3.1 \times 10^{-2}$ &
  $2.9 \times 10^{-4}$  &
  $3.6 \times 10^{-4}$ \\
$13$ &  &
  $10^{-2}$ & $6.1 \times 10^{-3} $
   \\
$23$ & & $6.1 \times 10^{-3}$ & $6.1 \times 10^{-3}$ \\
\noalign{\smallskip}\hline
\end{tabular}
}
\end{table}
Numerical values are listed in Table \ref{tab:ccbufb},
corresponding to universal sfermion masses.

\section{Low-energy phenomena}
\label{sec:pheno}
Each of the flavour-violating parameters $ (\delta^{\tilde f}_{ij})_{AB} $
changes flavour by one unit. Hence generically, they mediate
flavour-changing weak decays of $B$, $D$, $K$
mesons as well as charged leptons, such as $B \to X_s \gamma$
or $\tau \to \mu \gamma$, at linear order and $\Delta F=2$ processes,
namely particle-antiparticle mixing of neutral mesons,
at quadratic order. 
Such processes provide, at present, stringent constraints on
supersymmetric parameters, and eventually discovery and ``measurement''
potential for supersymmetry.

Most observables receive contributions from both the Standard Model
and from supersymmetry. Schematically,
\begin{equation}
  {\cal A} = {\cal A}_{\rm SM} + {\cal A}_{\rm SUSY} .
\end{equation}
A decay rate provides a (schematic) constraint
\begin{eqnarray}     \label{eq:constrdec}
    \lefteqn{|{\cal A}_{\rm SUSY}|^2 + 2\,{\rm Re}\,{\cal A}_{\rm
               SUSY}^* {\cal A}_{\rm SM} }
\\ \qquad         &=& \Gamma_{\rm exp} (1 \pm \Delta^{(\rm exp)})
            - |{\cal A}_{\rm SM}|^2 (1 \pm \Delta^{(\rm SM)}) .
\end{eqnarray}
By this we mean that the left-hand side, which can be calculated in
principle from the parameters of the SM and the MSSM, is constrained
to lie within a range (right-hand side) determined by the experimental
range (or bound) on the decay rate and uncertainties on the
theoretical prediction of its pure SM contribution. 
In the case of meson-antimeson mixing in principle information
on the complex ${\cal A}_M \equiv {\cal A}(\bar M \to M)$ is
accessible (once phase conventions are fixed), i.e.
\begin{eqnarray}
    {\cal A}_{M,\rm SUSY} &\equiv& (C_M e^{i 2 \phi_M} - 1) {\cal A}_{\rm SM}
  \\   \nonumber
       &=& {\cal A}_{M, \rm exp} (1 \pm \Delta^{(\rm exp)})
           - {\cal A}_{M, \rm SM} (1 \pm \Delta^{(\rm SM)})
\end{eqnarray}
(here strictly speaking one has two independent equations, including
errors, for magnitude and complex phase).
The right-hand sides often involve a ``cancellation'': the flavour observables
measured so far are consistent with no SUSY contributions.
Hence it is important to improve the precision on $\Delta^{(\rm exp)}$ and
$\Delta^{(\rm SM)}$ as much as possible; depending on the mode this
is mainly a theory or an experimental challenge.

Conversely, for the SUSY contributions on the left-hand side the leading
dependence on SUSY parameters is enough.

\subsection{Theoretical framework}

Weak decays are often either loop-induced in the Standard Model or
receive important loop corrections. Moreover, they are multi-scale
problems, involving large logarithms of physical scales
$M_{\rm SUSY} \gg M_L$, where $M_L$ denotes a generic mass of the
initial- and final-state particles
involved ($B$, $D$, $\tau$, $\mu$, light mesons, leptons and
neutrinos). In view of the smallness of the latter compared
to the weak and SUSY scales, effective theories provide a framework
for separating (factorizing) the contributions from the weak scale
and above from low-energy QCD and QED effects. Apart from the huge
simplification in theoretical expressions for the observables,
effective theories are the appropriate tool to curb large logarithms
$\ln M_W/m_L$ or $\ln M_{\rm SUSY}/m_L$, and to separate
nonperturbative QCD effects into a limited number of
well-defined hadronic matrix elements.

\subsubsection{SUSY scale and decoupling}
Consider first the case $M_{\rm SUSY} \gg M_W$. In this case, integrating out
the superpartners one readily obtains a set of higher-dimensional
local operators built from SM fields (including the Higgs and weak
bosons), suppressed by powers of $M_{\rm SUSY}$. As is well known, the
unique operator at dimension 5 violates lepton number by two units and
is not generated from the MSSM. A complete set of operators up at dimension 
six has been given in \cite{Buchmuller:1985jz}. Without computing any
loops, this implies that SUSY contributions (or those from any other new
physics with a new large mass scale $M$) to decay amplitudes
decouple as
\begin{equation}
    \Delta \{BR, A_{\rm CP}, \dots \}_{\rm SUSY}
           \sim \frac{\delta^{\tilde f}_{ij}}{M^2} ,
\end{equation}
while effects in meson-antimeson mixing decouple like
\begin{equation}
    \Delta \{ \epsilon_K, \Delta M_{B,D,K}, S_{J/\psi K},
     S_{J/\psi \phi}, \dots \}_{\rm SUSY}
           \sim \frac{\left(\delta^{\tilde f}_{ij}\right)^2}{M^2} ,
\end{equation}
in agreement with the decoupling theorem. In particular, the
flavour-changing vertices of the $Z$ and the magnetic vertices of
the photon and the gluon decouple as $M^{-2}$, despite being
apparently of dimension 4 and 5, respectively.
Conversely, it means that experimental bounds on the $\delta$
parameters scale, in general, linearly with $M$ for mixing and
quadratically for decays. Note the difference to the vacuum-stability
constraints \eq{eq:ccbfirst}--\eq{eq:ufblast}.

In general, of course, one expects that at least some of the
superpartners are close to the weak scale on naturalness grounds, such
that in practice it is often necessary to integrate out sparticles
and the weak scale at the same time.

\subsubsection{Weak scale}
The appropriate tool to separate the heavy
scales from the low-energy QCD effects and curb large QCD (and QED)
logarithms is the effective weak Hamiltonian (see~\cite{Buras:1998raa} for a
review of the formalism). Integrating out the weak scale, one has
\begin{equation}
  {\cal A}(i \to f) = \langle f | {\cal H}_{\rm eff} | i \rangle ,
\end{equation}
where the weak Hamiltonian
\begin{equation}
  \label{eq:heff}
  {\cal H}_{\rm eff} = \sum_k C_k(\mu) Q_k(\mu)
\end{equation}
is given in terms of local operators $Q_k$
constructed from the ``light'' SM fields
($u,d,s,c,b$, $\tau, \mu, e$, photon, gluons, and neutrinos) and of
Wilson coefficients $C_k$ encapsulating the effects of the
weak interactions and the sparticles, and
$\mu$ is a renormalization (factorization) scale. Radiative
corrections from above the scale $\mu$ are contained in the $C_k(\mu)$.

The Wilson coefficients are evolved to lower scales according to
RG equations
\begin{equation}
      \mu \frac{{\rm d}}{{\rm d} \mu} C_k(\mu)
           = \gamma_{lk}(\alpha_s(\mu), \alpha)\, C_l(\mu) ,
\end{equation}
where $(\gamma_{lk})$ are perturbative, process-independent anomalous-dimension
matrices.

\subsubsection{Lower scales}
In a purely leptonic decay such as $\tau \to \mu \gamma$, the matrix
element of the  weak hamiltonian can be simply calculated in
perturbation theory. (In fact, in this case the use of the weak
Hamiltonian is not very essential due to the absence of large
radiative corrections.) For the large amount of data that
involve hadrons, one has only
\begin{equation}
    {\cal A}(i \to f) = \sum_k C_k(\mu) \langle f | Q_k(\mu) | i
    \rangle \equiv \sum_k C_k(\mu) B_k(i,f) ,
\end{equation}
where $\mu$ is optimally chosen of order of the mass of $i$. The
hadronic matrix elements $\langle f | Q_k(\mu) | i \rangle$
are usually nonperturbative and only calculable in some cases.
The latter include matrix elements for meson-antimeson mixing,
which can be obtained using numerical lattice QCD methods.
Other methods include QCD sum rules based on the operator product
expansions (for inclusive and some exclusive $B$, as well
as hadronic $\tau$ decays) and collinear expansions (for some
exclusive $B$ decays), chiral perturbation theory in $K$ decays,
and the use of approximate flavour symmetries of QCD
to reduce the number of independent hadronic matrix elements;
all of these have systematics controlling which is a theoretical
challenge.

\subsection{$K^0 -\bar K^0$, $B^0-\bar B^0$, $B_s -\bar B_s$, and
  $D^0-\bar D^0$ mixing}
Meson mixings are $\Delta F=2$ processes.
At one loop, the effective $\Delta F=2$ hamiltonian to meson-antimeson
oscillations is solely due to box diagrams.
Complete operator bases have been given
in~\cite{Gabbiani:1996hi,Buras:2000if}.
For $\Delta B= \Delta S=2$ transitions ($B_s-\bar B_s$ mixing),
one choice consists of the five operators
\begin{eqnarray}
Q_1 &=& (\bar s_L^a \gamma_\mu b_L^a) (\bar s_L^b \gamma^\mu  b_L^b), \\
Q_2 &=& (\bar s_R^a b_L^a) (\bar s_R^b b_L^b), \\
Q_3 &=& (\bar s_R^a b_L^b) (\bar s_R^b b_L^a), \\
Q_4 &=& (\bar s_R^a b_L^a) (\bar s_L^b b_R^b), \\
Q_5 &=& (\bar s_R^a b_L^b) (\bar s_L^b b_R^a)
\end{eqnarray}
($a,b$ colour indices), plus operators $\tilde Q_{1,2,3}$ obtained
by flipping the chiralities of all fermions in $\tilde Q_{1,2,3}$.
The operator basis for $B_d-\bar B_d$, $D^0 - \bar D^0$,
and $K^0-\bar K^0$ mixing are identical up to obvious substitutions
of quark flavours
(in the case of  $K^0-\bar K^0$ and $D^0-\bar D^0$ 
mixing, there are also
sizable ``long-distance'' contributions which cannot be written in
terms of local four-quark operators at the weak scale).

\begin{figure}
\vskip0.5cm
\hspace*{1cm}
\includegraphics[width=7cm,angle=0]{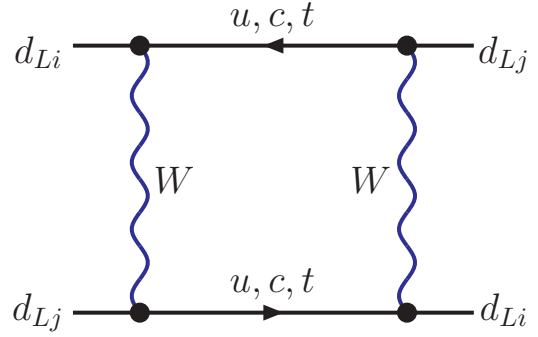}
\vskip0.9cm
\caption{SM diagram for neutral meson-antimeson mixing. 
(Diagrams including Goldstone bosons in $R_\xi$ gauge not shown.)
\label{fig:df2sm}       
}
\end{figure}
Only $Q_1$ is generated in the SM (to excellent approximation), following
from $W-t$ boxes (Fig. \ref{fig:df2sm}.) This results in
\begin{equation}
   C_1^{\rm SM} = \frac{G_F^2 M_W^2}{16 \pi^2} (V_{tb} V_{ts}^*)^2 4\, S(x_t) ,
\end{equation}
where $S$ \cite{Inami:1980fz} is listed in appendix \ref{app:loopfun}.
SM NLO QCD corrections are reviewed in \cite{Buras:1998raa}.

Supersymmetric contributions have been computed in
\cite{Gabbiani:1996hi,Gerard:1984bg,Bertolini:1987cw,Gabbiani:1988rb,Hagelin:1992tc,Bertolini:1990if,Urban:1997gw,Feng:2000kg,Ciuchini:2006dw,Gabrielli:1995bd}.
Since each $\delta$ changes flavour by one unit, the leading
contributions are of second order in these parameters.
\begin{figure*}
\vskip0.5cm
\hspace*{1cm}
\includegraphics[width=16cm,angle=0]{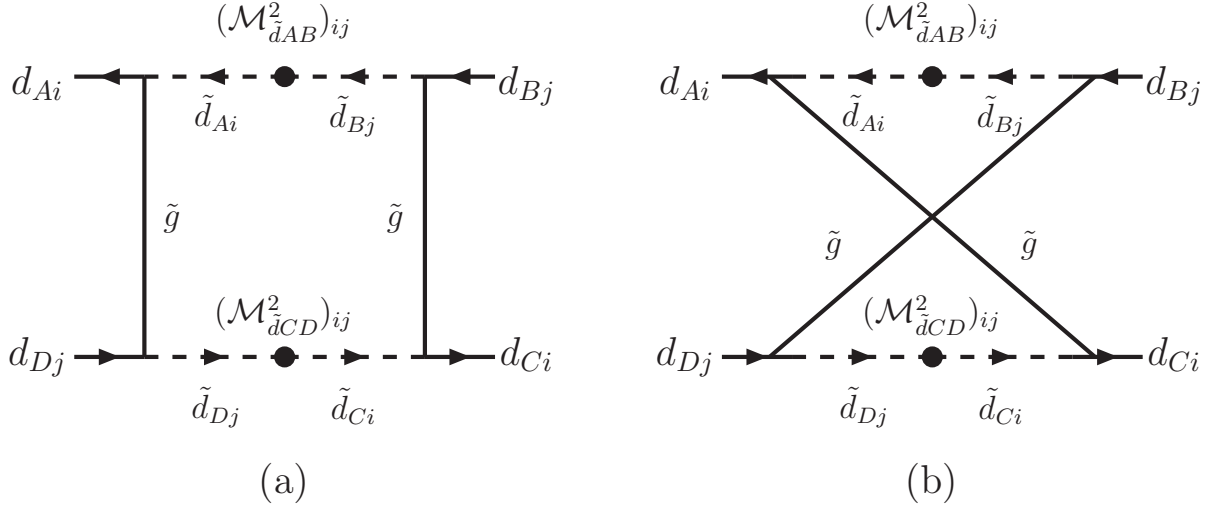}
\vskip0.9cm
\caption{Diagrams for meson-antimeson mixing. $A,B,C,D$ denote
  chiralities of the quarks (and squarks). The blobs are
flavour-changing ``mass insertions''.
\label{fig:df2mia}       
}
\end{figure*}
The simplest way to obtain the second-order terms is to work in the
``mass-insertion approximation'', where the off-diagonal
sfermion-mass-matrix elements 
are treated as perturbations (Fig. \ref{fig:df2mia}).
 For instance, for two LL mass insertions,
diagram \ref{fig:df2mia} (a) (to zeroth order in external momenta,
and neglecting mass differences between the squarks in the loop)
is proportional to
\begin{eqnarray}
&&   \int {\rm d}^4k
     \frac{k^2 (\cM^2_{\tilde dLL})_{sb}^2}{(k^2-m_{\tilde g}^2)^2(k^2-m_{\tilde q}^2)^4}
  \nonumber \\
&=& \frac{(\delta^{\tilde d}_{sb} )_{LL}^2}{6}
    \frac{(m_{\tilde q}^2)^2{\rm d}^2}{({\rm d} m^2_{\tilde q})^2}
      \int {\rm d}^4 k \frac{k^2}{(k^2-m_{\tilde g}^2)^2(k^2-m_{\tilde
          q}^2)^2} .
\end{eqnarray}
The full result for the gluino-squark contributions reads~\cite{Gabbiani:1996hi}
\begin{eqnarray}
   C_1 &=& - \epsilon   \label{eq:df2mia1st}
      [24 x f_6(x)  + 66 \tilde f_6(x)] \, (\delta^\td_{sb})_{LL}^2 , \\
   \tilde C_1 &=& - \epsilon  
      [24 x f_6(x)  + 66 \tilde f_6(x)] \, (\delta^\td_{sb})_{RR}^2 , \\
   C_2 &=& -\epsilon\, 204 x f_6(x) \, (\delta^\td_{sb})_{RL}^{2}, \\
   \tilde C_2 &=& - \epsilon\, 204 x f_6(x) \, (\delta^\td_{sb})_{LR}^2, \\
   C_3 &=& \epsilon\, 36 x f_6(x) \, (\delta^\td_{sb})_{RL}^{2}, \\
   \tilde C_3 &=& \epsilon\, 36 x f_6(x) \, (\delta^\td_{sb})_{LR}^2, \\
   C_4 &=& - \epsilon [504 x f_6(x) - 72 \tilde f_6(x)] \,
               (\delta^\td_{sb})_{LL} (\delta^\td_{sb})_{RR} 
 \nonumber \\ && + \epsilon\, 132 \tilde f_6(x) \,
      (\delta^d_{sb})_{LR} (\delta^\td_{sb})_{RL} , \\
   C_5 &=& - \epsilon [24 x f_6(x) + 120 \tilde f_6(x)] \,
                     (\delta^\td_{sb})_{LL} (\delta^\td_{sb})_{RR}
 \nonumber \\ && + \epsilon\, 180 \tilde f_6(x) \,
      (\delta^\td_{sb})_{LR} (\delta^\td_{sb})_{RL} .
         \label{eq:df2mialast}
\end{eqnarray}
Here $(\delta^\td_{ij})_{RL} \equiv (\delta^\td_{ji})_{LR}^*$,
$\epsilon = \alpha_s^2/(216\,m_{\tilde q}^2)$ ,
 $x = m_{\tilde g}^2/m_{\tilde q}^2$, and $f_6(x)$, $\tilde f_6(x)$
are dimensionless loop functions (appendix \ref{app:loopfun})

There are also chargino-up-squark contributions. These can be
competitive with the gluino-squark contributions if the charginos
are lighter than the gluinos, as tends to be the case in GUT
scenarios. There are always ``minimally flavour-violating'' contributions,
which are proportional to the same CKM factors as the SM
contributions.
Of interest here are the additional
contributions due to nonvanishing $\delta^u$ parameters.
Neglecting terms suppressed by small CKM elements or small
Yukawa couplings, only $C_1$ receives a contribution \cite{Colangelo:1998pm}
\begin{eqnarray}
   C_1 &=& - \frac{G_F \alpha}{\sqrt{2} \pi \sin^2 \theta_W}
   \frac{M_W^2}{m_{\tilde q}^2} \nonumber \\
  && \times\frac{1}{20} \left[
   ( [\delta^\tu_{ij})_{LL}]^2
   - \frac{2}{3} (\delta^\tu_{ij})_{LL} (\delta^\tu_{it})_{LR}
   (\delta^\tu_{jt})_{LR}^* \right.
\nonumber \\
  && \left. \qquad  + \frac{1}{7} [(\delta^\tu_{it})_{LR}
    (\delta^\tu_{jt})_{LR}^*]^2 \right] .
\end{eqnarray}
Note that the chargino contributions involve either a $LL$ mass
insertion or a double $LR$ one on each squark line; for the latter,
only those involving a stop can be relevant according to 
Table \ref{tab:ccbufb}. (For $B-\bar B$ mixing, there may
be additional operators \cite{Gabrielli:2002fr}.)

If $\tan\beta$ is large, there are in principle also
terms proportional to $y_b$ that could be important. In that
case, however, Higgs double-penguin diagrams are often dominant
and require a modified treatment
\cite{Buras:2001mb,Isidori:2001fv,Buras:2002wq,Buras:2002vd}.

\subsubsection{$K-\bar K$ mixing  and constraints on $\delta$'s}
$K-\bar K$ oscillations proved their discovery potential in estimating the
charm quark mass before its observation \cite{Gaillard:1974hs}, as well
as in the discovery of (indirect) CP violation
\cite{Christenson:1964fg},
later giving information on the CP-violating phase in the CKM matrix.
The possibility of large SUSY contribution was recognized early
on \cite{Dimopoulos:1981zb,Ellis:1981ts,Barbieri:1981gn,Duncan:1983iq,Donoghue:1983mx}, and $\Delta M_K$ and $\epsilon_K$ still provide the
strongest FCNC constraints on the MSSM parameters.
The mass difference $\Delta M_K$ and the CP-violating parameter
$\epsilon_K$ follow from the effective $\Delta F=2$ Hamiltonian,
\begin{eqnarray}
  \Delta M_K
&\propto& 
  2 \sum_i B_i\, {\rm   Re}\, C_i , \\
  \epsilon_K &\propto& \frac{e^{i \pi/4}}{\sqrt{2} \Delta M_K}
    \sum_i B_i\, {\rm Im}\, C_i,
\end{eqnarray}
where $B_i \equiv \langle K | Q_i | \bar K \rangle$. The hadronic matrix
elements $B_i$ contain low-energy QCD effects and require
nonperturbative methods such as (numerical) lattice QCD,
see e.g.~\cite{Donini:1999nn,Babich:2006bh,Becirevic:2001xt}.\footnote{
Usually, the hadronic matrix elements are normalized to their
values obtained from PCAC in "vacuum-insertion approximation".
This normalization is included in the $B_i$ here. }
Moreover, $\Delta M_K$ is afflicted by long-distance contributions
which are believed to be not much larger than the SM short-distance
contribution but are difficult to estimate.
Nevertheless, in view of the strong CKM suppression of the SM
contribution, even a rough estimate of the
$B_i$ translates into strong constraints on $s\to d$ flavour violation
parameters. The procedure is as follows \cite{Gabbiani:1996hi}:
\begin{itemize}
   \item Write out the expression for the observable (here, $\epsilon_K$ or
   $\Delta M_K$) as linear combination of (products of)
   $\delta$-parameters, inserting estimates of the hadronic matrix elements.
   \item Require that each term at most saturates the experimental
   result.
\end{itemize}
This is clearly a crude procedure. Typically, there will be
some degree of constructive or destructive interference between
different terms and with the SM contribution. However, in view of
the large number of parameters this is a reasonable approach:
Even with a more precise determination of the Standard Model
contribution (including the relevant hadronic matrix elements and
nonlocal contributions),
the possible presence of cancellations prohibits a significant improvement
of this type of constraint without further assumptions.

\begin{table}
\caption{Constraints from $K-\bar K$ mixing on $\delta$
parameters \cite{Gabbiani:1996hi,Misiak:1997ei} for
$m_{\tilde q}= 500$ GeV and $x= m_{\tilde g}^2/m_{\tilde q}^2 = 1$.
Bounds vary by less than one order of magnitude in the range
$x \in [0.3, 4.0]$. For the bounds on $LR$ mass insertions,
$(\delta^\td_{ds})_{LR} \sim (\delta^\td_{ds})_{RL}$ has been
assumed \cite{Gabbiani:1996hi}.
\label{tab:miaK}     }  
\centerline{
\begin{tabular}{lc}
\hline\noalign{\smallskip}
 Quantity & upper bound \\
\noalign{\smallskip}\hline\noalign{\smallskip}
$\sqrt{|{\rm Re} (\delta^\td_{ds})^2_{LL}}|$ & $4.0 \times 10^{-2}$ \\[2mm]
$\sqrt{|{\rm Re} (\delta^\td_{ds})^2_{RR}}|$ & $4.0 \times 10^{-2}$ \\[2mm]
$\sqrt{|{\rm Re} (\delta^\td_{ds})^2_{LR}|}$ & $4.4 \times 10^{-3}$ \\[2mm]
$\sqrt{|{\rm Re} (\delta^\td_{ds})_{LL}(\delta^\td_{ds})_{RR} |}$ & $2.8 \times 10^{-3}$ \\[2mm]
$\sqrt{|{\rm Im} (\delta^\td_{ds})^2_{LL}}|$ & $3.2 \times 10^{-3}$ \\[2mm]
$\sqrt{|{\rm Im} (\delta^\td_{ds})^2_{RR}}|$ & $3.2 \times 10^{-3}$ \\[2mm]
$\sqrt{|{\rm Im} (\delta^\td_{ds})^2_{LR}|}$ & $3.5 \times 10^{-4}$ \\[2mm]
$\sqrt{|{\rm Im} (\delta^\td_{ds})_{LL}(\delta^\td_{ds})_{RR} |}$ & $2.2 \times 10^{-4}$ \\[2mm]
\noalign{\smallskip}\hline
\end{tabular}
}
\end{table}
For the case of $K-\bar K$ mixing, this leads to the bounds
\cite{Gabbiani:1996hi,Misiak:1997ei}
shown in Table \ref{tab:miaK}
(assuming gluino-squark dominance and absence of cancellations). This
is one case of the ``SUSY flavour problem''.
The constraints become weaker as the SUSY scale is increased,
scaling like $M$, as is required by the decoupling properties
discussed above and is evident from~\eq{eq:df2mia1st}--\eq{eq:df2mialast}.
At any rate, this ``problem'' looks less severe when considering that
the corresponding CKM factor $V_{td}^* V_{ts}= {\cal O}(10^{-4})$ is
also much smaller than its `generic' value ${\cal O}(1)$, and that
the $LR$ $\delta$ parameters are ${\cal O}(v/M)$;
moreover, already the vacuum-stability bounds considered
in Subsection \ref{sec:ccbufb} show that the squark mass matrices
(in particular, the left-right elements deriving from the $T$-parameters)
must be far from generic. As previously noted, the problem
is completely removed, for instance, in gauge mediation, which forms
a special case of the framework of minimal flavour
violation~\cite{Hall:1990ac,Buras:2000dm,Buras:2000qz,D'Ambrosio:2002ex}.

Slightly more stringent constraints on the mass insertions have been obtained in
\cite{Ciuchini:1998ix}, which uses better hadronic matrix elements
from a lattice calculation and imposes the (stronger) requirement
that for $\Delta m_K$ and $\epsilon_K$, the SUSY contribution from any
single biproduct of mass insertions does not exceed the difference
between the (short-distance) SM contribution and the measured value.
One has
\begin{eqnarray}
|(\delta^\td_{ds})_{LL}|, |(\delta^\td_{ds})_{RR}| &<& {\cal O}(10^{-2}),
    \label{eq:deltaLLbound}
\\
|(\delta^\td_{ds})_{LR}|, |(\delta^\td_{ds})_{RL}| &<& {\cal O}(10^{-3}), \\
|(\delta^\td_{ds})_{LL} \cdot (\delta^\td_{ds})_{RR}| &<& {\cal O}(10^{-7})
    \label{eq:deltaLLRRbound}
\end{eqnarray}
for $M \sim m_{\tilde q} \sim m_{\tilde g} \sim 500$ GeV~\cite{Ciuchini:1998ix}.
In particular, the bounds on products of $LL$ and $RR$ insertions
are strengthened. In that paper, specifics on the procedure and
more detailed  numerical results can be found.

\subsubsection{$D - \bar D$ mixing}
With the observation of $D-\bar D$ oscillations in 2007 at the
$B$-factories~\cite{Aubert:2007wf,Staric:2007dt,Abe:2007rd},
particle-antiparticle mixing has now been established in all four
of the neutral meson-antimeson systems.  Unfortunately it is
very difficult to quantify the SM contribution to the mixing
amplitude, as it
is completely long-distance dominated and rather uncertain.
Hence possible new-physics short-distance contributions are only
limited by the experimental upper bound (barring cancellations with the
unknown SM contribution), which has now been replaced and improved by a
concrete measurement. The situation is very different for possible
CP violation: as the  mixing amplitude must have a negligible weak
phase in the SM (being due to only the first two generations),
mixing-induced CP violation in $D$ decays would clearly signal non-SM physics.
Implications of the improved bound for new physics,
including supersymmetry, have ben discussed
in~\cite{Ciuchini:2007cw,Nir:2007ac,Golowich:2007ka,Fajfer:2007dy}.
In particular, since $(\delta^u_{12})_{LL}$ and $(\delta^d_{12})_{LL}$
mass insertions are related to each other and the mass splitting
between $\tilde d_L$ and $\tilde s_L$
by the Cabibbo angle, it appears less likely
that squark and quark masses are merely aligned; rather a near
degeneracy of at least the left-chiral squarks of the first two
generations is necessary \cite{Ciuchini:2007cw,Nir:2007ac,Golowich:2007ka}.

\begin{table}
\caption{Constraints from $D^0-\bar D^0$ mixing on $\delta^u$
parameters. Here we have rescaled the results from \cite{Gabbiani:1996hi}
(for $m_{\tilde q}= 500$ GeV and $x= m_{\tilde g}^2/m_{\tilde q}^2 =
1$) to account for the replacement of the old experimental upper bound
$\Delta M_D < 1.32 \times 10^{-10}$ MeV used there by the current
one-sigma upper bound $\Delta M_D < 1.99 \times 10^{-10}$ MeV
\cite{Amsler:2008zz}, otherwise the remarks concerning Table
\ref{tab:miaK} apply.
\label{tab:miaD}     }  
\centerline{
\begin{tabular}{lc}
\hline\noalign{\smallskip}
 Quantity & upper bound \\
\noalign{\smallskip}\hline\noalign{\smallskip}
$\sqrt{|{\rm Re} (\delta^\tu_{uc})^2_{LL}}|$ & $3.9 \times 10^{-2}$ \\[2mm]
$\sqrt{|{\rm Re} (\delta^\tu_{ud})^2_{RR}}|$ & $3.9 \times 10^{-2}$ \\[2mm]
$\sqrt{|{\rm Re} (\delta^\tu_{uc})^2_{LR}|}$ & $1.20 \times 10^{-2}$ \\[2mm]
$\sqrt{|{\rm Re} (\delta^\tu_{uc})_{LL}(\delta^u_{uc})_{RR} |}$ & $6.6 \times 10^{-3}$ \\[2mm]
\noalign{\smallskip}\hline
\end{tabular}
}
\end{table}
Considering gluino-squark box diagrams and following the same
procedure as outlined for $K^0 - \bar K^0$ mixing above,
the constraints on the up-squark mass matrices listed
in Table \ref{tab:miaD} result.

\subsubsection{$B_d-\bar B_d$ and $B_s-\bar B_s$ mixing} \label{sec:bbmix}
Here the mixing amplitudes
\begin{equation}
  {\cal A}(\bar B_q \to B_q) \propto M_{12}^q - \frac{i}{2} \Gamma_{12}^q 
\end{equation}
($q=d,s$)
are completely short-distance dominated. Hence the theoretical
expression
\begin{equation}
  \Delta M_{B_q} \propto |M_{12}^q| \sim f_{B_q}^2 | \sum B_i R_i C_i | ,
\end{equation}
where $f_{B_q}$ denotes a decay constant and $R_i$ a known (up
to mild uncertainties) factor defined by the vacuum insertion approximation
($\langle B_q | Q_i | \bar B_q \rangle \equiv f_{B_q} R_i B_i$)
can be directly compared to the experimental
values~\cite{Barberio:2007cr,Abulencia:2006ze}
\begin{eqnarray}
    \Delta M_{B_q} &=& (0.507 \pm 0.005)\, {\rm ps}^{-1},  \\
    \Delta M_{B_s} &=& (17.77 \pm 0.10 \pm 0.07)\, {\rm ps}^{-1} .
\end{eqnarray}
In both cases, the theory error is fully dominated by $f_{B_q}$. For instance,
$\Delta M_{B_s}^{\rm SM} \approx (16 \dots 27)\, {\rm ps}^{-1}$ is achievable
depending on which value of $f_{B_s}$
\cite{utfit,ckmfit,tantalo,DellaMorte:2007ny} is used.
 Conversely, the residual ``bag factors''
$B_i$ have much smaller uncertainties.
The SM prediction is consistent with the experimental value.
However, the theoretical range clearly allows for several tens of
percent of new-physics contribution, in particular if it
is mostly imaginary (i.e. CP violating). See also
Section \ref{sec:bsmhints} below.
Combined with the fact that the remaining
(perturbative, non-CKM) uncertainties are at the 1--2 percent level,
this underlines the importance of the ongoing efforts to obtain
these nonperturbative parameters on the lattice with a high precision.

On the other hand, the weak phase $\phi_d = \arg M^{d}_{12}$ governs
mixing-decay interference, and can be extracted cleanly from the
time-dependent CP asymmetry in
$B \to J/\psi K_S$ decay (with theoretical uncertainties
of order $1-2$ \%), giving~\cite{Barberio:2007cr}
\begin{equation}
    \sin \phi_d = 0.680 \pm 0.025 .
\end{equation}
In the SM, $\phi_d = 2 \beta$, but this does not hold in the presence
of new flavour violation. Further information on the mixing phase
(more precisely, on $\arg (-M_{12}/\Gamma_{12})$, which in the MSSM
and other new-physics scenarios without new light degrees
of freedom can be related to $\arg M_{12}$
using a robust theoretical framework, see \cite{Lenz:2006hd} for a recent
theoretical account) can be obtained from flavour-specific
$B_d$ and $B_s$ decays, notably the so-called semileptonic CP asymmetries.

One of the most exciting developments in the past year has been
the first measurement of the mixing phase
$\phi_s$\footnote{We caution the reader that the literature abounds
with conventions and meanings of the symbol $\phi_s$ and
similar phases used in parameterizing new-physics effects.
Here, $\phi_s \equiv \arg M_{12}^s$ in the standard CKM parameterization
[Ref. \cite{Lenz:2006hd} uses $\phi_s$ to denote the
(CKM-parameterization-invariant) phase
$\arg(-M_{12}^s/\Gamma_{12}^s)$]. Furthermore, in the standard
parameterization one has $\beta \approx - \arg V_{td}$, however
$\beta_s \approx + \arg (-V_{ts})$.
}
in the $B_s$ system
by the D0 and CDF collaborations at Fermilab from an analogous
time-dependent CP asymmetry in  $B_s \to J/\psi\, \phi$ decays.
The current $68$ \% CL average of CDF and D0 data reads \cite{phisTevatronIchep}
\begin{equation}
    -65^\circ < \phi_s < -27^\circ \vee -152^\circ < \phi_s < -113^\circ .
\end{equation}
The statistical accuracy will be further reduced by CDF and D0 and
by LHCb, with a goal of about $1^\circ$ by 2013
\cite{LibbyFlavLHC}.
As in the SM $\phi_s \approx -2 \beta_s = -2.2^\circ \pm 0.6^\circ$
\cite{Lenz:2006hd}, such a  mixing-induced
asymmetry if confirmed would be an unambiguous, theoretically
clean signal of not only new but nonminimally flavoured physics
(see Section \ref{sec:bsmhints} below).

\begin{table}
\caption{Constraints from $B_d^0-\bar B_d^0$ \cite{Gabbiani:1996hi}
and $B_s^0-\bar B_s^0$ mixing on $\delta^d$ parameters. 
For the latter, we have rescaled the $B_d^0$ bounds
to account for the larger mass difference
$\Delta M_{B_s} = (117.0 \pm 0.8)\times 10^{-10}$ MeV
(as opposed to $\Delta M_{B_d} = (3.337 \pm 0.033) \times 10^{-10}$
MeV) \cite{Amsler:2008zz} and the non-unit ratio of the decay constants,
which we take to be
$f_{B_s}/f_{B_d} = 1.2$ in agreement with lattice-QCD results
(see \cite{DellaMorte:2007ny} for a review).
Otherwise the remarks concerning Table
\ref{tab:miaK} apply.
\label{tab:miaB}     }  
\centerline{
\begin{tabular}{lc}
\hline\noalign{\smallskip}
 Quantity & upper bound \\
\noalign{\smallskip}\hline\noalign{\smallskip}
$\sqrt{|{\rm Re} (\delta^\td_{db})^2_{LL}}|$ & $9.8 \times 10^{-2}$ \\[2mm]
$\sqrt{|{\rm Re} (\delta^\td_{db})^2_{RR}}|$ & $9.8 \times 10^{-2}$ \\[2mm]
$\sqrt{|{\rm Re} (\delta^\td_{db})^2_{LR}|}$ & $3.3 \times 10^{-2}$ \\[2mm]
$\sqrt{|{\rm Re} (\delta^\td_{db})_{LL}(\delta^\td_{db})_{RR} |}$ & $1.8 \times 10^{-2}$ \\[2mm]
$\sqrt{|{\rm Re} (\delta^\td_{sb})^2_{LL}|}$ & $4.8 \times 10^{-1}$ \\[2mm]
$\sqrt{|{\rm Re} (\delta^\td_{sb})^2_{RR}|}$ & $4.8 \times 10^{-1}$ \\[2mm]
$\sqrt{|{\rm Re} (\delta^\td_{sb})^2_{LR}|}$ & $1.62 \times 10^{-2}$ \\[2mm]
$\sqrt{|{\rm Re} (\delta^\td_{sb})_{LL}(\delta^\td_{sb})_{RR} |}$ & $8.9 \times 10^{-2}$ \\[2mm]
\noalign{\smallskip}\hline
\end{tabular}
}
\end{table}
Several authors have considered the impact of the oscillation measurements
on the general MSSM.  Constraints on $\delta$'s analogous to those from
the neutral $K$ and $D$ systems follow follow from $B_d- \bar B_d$
(see \cite{Gabbiani:1996hi,Becirevic:2001jj}) and from $B_s-\bar B_s$
(see \cite{Foster:2006ze}, which treats the large $\tan\beta$
case where also Higgs double-penguin diagrams can make important
contributions, and \cite{Ball:2006xx,Ciuchini:2006dx}) mixing data. Bounds
following from the mass differences according to the prescription
of \cite{Gabbiani:1996hi} are listed in Table \ref{tab:miaB}.
See also  Subsection~\ref{sec:global}, Section~\ref{sec:susygut}
in the context of SUSY GUTs, and Subsection \ref{sec:bsmphis}
for the impact of possible CP violation in the $B_s$ system.

\subsection{Flavour-changing decays: higgs penguins}
For weak decays, only one quark flavour needs to be changed, which implies
contribution from penguin graphs. (Of course, there are also box
graphs.) Higgs penguins generating operators such as
$$
\sum_q y_q (\bar s_{L} b_{R}) (\bar q_L q_R)
$$
are negligible in the SM, due to the smallness of down-type
Yukawa coulings. This continues to hold in the MSSM for
small $\tan\beta$. As previously mentioned, at
large $\tan\beta$, depending on the Higgs sector this may change.
There is a peculiar phenomenology already for minimal flavour
violation, see e.g.\
\cite{Hamzaoui:1998nu,Choudhury:1998ze,Babu:1999hn,Isidori:2001fv,Buras:2002wq,Buras:2002vd,Carena:2000uj,Dedes:2002er,Isidori:2006jh,Isidori:2006pk,Freitas:2007dp,Ellis:2007kb,Trine:2007ma,GJNT}
and \cite{Isidori:2007ed} for a review;
the possible relevance of double Higgs penguins for $B-\bar B$ mixing
has already been mentioned.
For detailed studies of large-$\tan\beta$ effects
beyond minimal flavour violation we refer
to~\cite{Foster:2006ze,Foster:2004vp,Foster:2005wb,Foster:2005kb}.

\subsection{Electroweak penguins}

Photon and $Z$ penguins contribute to operators (shown here for
$b\to s$ transitions)
\begin{eqnarray}    \label{eq:ewpopsfirst}
  Q_{7\gamma} &=& - \frac{e\, m_b}{4 \pi^2}
        \bar s_{L}\, \sigma^{\mu \nu} b_{R} F_{\mu\nu}, \\
  Q_{9V}\,  &=&
      (\bar s_{L} \gamma^\mu b_{L}) (\bar l \gamma_\mu \bar l), \\
  Q_{10A} &=&
      (\bar s_{L} \gamma^\mu b_{L}) (\bar l \gamma_\mu \gamma_5 \bar l), \\
  Q_{9,10} &=&
    \sum_q \frac{3 e_q}{2} (\bar s_{L}\gamma^\mu b_{L}) (\bar q_L
    \gamma_\mu q_L) , \\
  Q_{7,8} &=&
      \sum_q \frac{3 e_q}{2} (\bar s_{L} \gamma^\mu b_{L})
     (\bar q_R \gamma_\mu q_R) ,  \label{eq:ewpopslast}
\end{eqnarray}
as well as their mirror images $Q_i'$ (obtained by flipping the
chiralities of all fermions). $F_{\mu \nu}$ is the electromagnetic
field strength. We have suppressed colour structures differentiating
between pairs of operators as indicated on the lhs of each line.
These operators also have analogs contributing to
lepton-flavour-violating $l_i \to l_j$ transitions.

In the SM, only the electroweak penguin (EWP) operators shown
in \eq{eq:ewpopsfirst}--\eq{eq:ewpopslast} receive significant
contributions, which are negligible for the ``opposite-chirality''
operators $Q_i'$, due to the V-A structure of the charged currents,
which are the only source of flavour violation. (This continues to
hold in the MSSM for minimal flavour violation.)
Identifying contibutions to some process from any of the primed
operators would constitute a clear signal of new physics.
Tree-level $W$-boson exchange also generates operators
\begin{eqnarray}
  Q^u_{1,2} &=& (\bar s_L \gamma^\mu u_L) (\bar u_L \gamma_\mu b_L), \\
  Q^c_{1,2} &=& (\bar s_L \gamma^\mu c_L) (\bar c_L \gamma_\mu b_L) ,
\end{eqnarray}
which do not receive comparable contributions in the MSSM.

All penguin operators except the magnetic ones also receive
contributions from box diagrams;
moreover, boxes can contribute to generic four-fermion operators
$$Q = (\bar s_{L,R}\, b_{L,R}) (\bar q_{L,R}\, q_{L,R}) .$$
(Here we have suppressed the several possibilities for the Dirac
structure.)
Note that, in general, boxes are not suppressed with
respect to penguin diagrams~\cite{Gabbiani:1996hi,Gabrielli:1995bd},
as all SUSY penguins as well as boxes decouple
as $M_W^2/M^2$ for $M$ large, as discussed before.

\subsubsection{Inclusive $B \to X_{s} \gamma$}
Although loop-induced already in the Standard Model,
the inclusive decay $B \to X_{s} \gamma$ has a relatively large
branching ratio and is well measured, with a world
average~\cite{Barberio:2007cr} of
\begin{equation}
   BR(\bar B \to X_s \gamma)_{\rm exp} = (3.52 \pm
   0.23 \pm 0.09) \times 10^{-4}
\end{equation}
(with a 1.6 GeV lower cut on the photon energy).
In the SM, the decay amplitude receives its dominant contributions from
$W-t$ loops entering through the magnetic operator $Q_{7\gamma}$ and $W-c$
loops entering through loop contractions of the tree operators
$Q^c_{1,2}$. Both contributions are of comparable size and opposite in
sign. Other operators are subdominant. The corresponding
state-of-the-art NNLO-QCD prediction reads~\cite{Misiak:2006zs}
(see~\cite{Haisch:2007ic} for a review and discussion of uncertainties)
\begin{equation}
   BR(\bar B \to X_s \gamma)_{\rm SM} = (3.15 \pm 0.23) \times 10^{-4},
\end{equation}
slightly more than $1 \sigma$ below the experiment.
Supersymmetric effects have been investigated
thoroughly in the literature both in minimal
flavour violation~\cite{D'Ambrosio:2002ex,Bertolini:1990if,Buras:2002vd,Carena:2000uj,Bertolini:1986tg,Bobeth:1999ww,Ciuchini:1998xy,Borzumati:2003rr,Degrassi:2006eh,Degrassi:2000qf}
and beyond~\cite{Gabbiani:1996hi,Gabbiani:1988rb,Hagelin:1992tc,Borzumati:1999qt,Besmer:2001cj,Ciuchini:2002uv}. 
They enter mainly through $Q_{7\gamma}$ and the chromomagnetic
operator $Q_{8g}$ (defined in \eq{eq:chromomag} below), which has a large mixing
with $Q_{7\gamma}$ under renormalization This results in a
large sensitivity to the parameters $(\delta^\td_{sb})_{LR}$
and $(\delta^\tu_{ct})_{LR}$, which allows to constrain these parameters,
for which we refer to the cited literature and to \cite{Misiak:1997ei}.
Complete expressions for the magnetic operators that are
valid for general flavour violation can be found
in~\cite{Borzumati:1999qt}. See also \cite{Cho:1996we}.
Contributions via $Q'_{7\gamma}$ do not interfere with the SM
contribution in an inclusive process due to the opposite chirality of
the produced $s$-quark, hence generally have a
small impact. The full one-loop SUSY contribution may
involve a cancellation between charged-Higgs-top loops, which are
always of the same sign as the SM piece, and squark-higgsino as well
as squark-gaugino loops, which (in general) carry an arbitrary complex phase.

\subsubsection{$Z$-penguins and rare $K$,$B$ decays}
\begin{figure*}
\vskip0.5cm
\hspace*{1cm}
\includegraphics[width=16cm,angle=0]{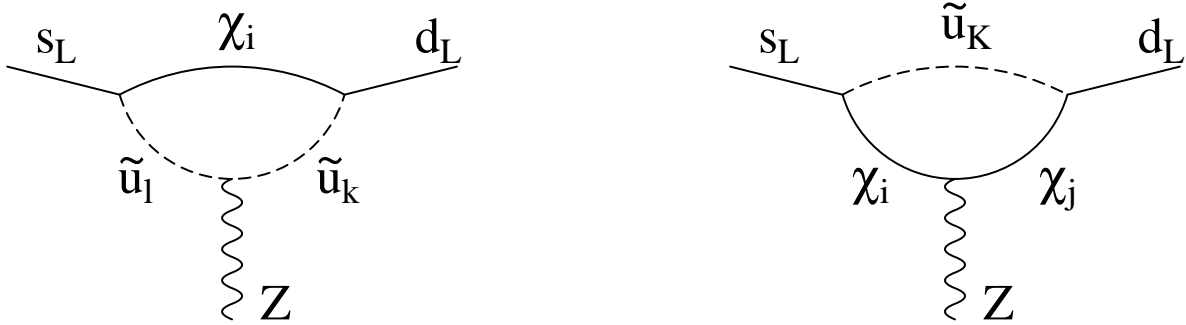}
\vskip0.9cm
\caption{Potentially large contributions to the $Z$-penguin (shown here
for $s\to d$ transitions \cite{Colangelo:1998pm})
\label{fig:zpengcharg}       
}
\end{figure*}
\begin{figure}
\includegraphics[width=0.45\textwidth,angle=0]{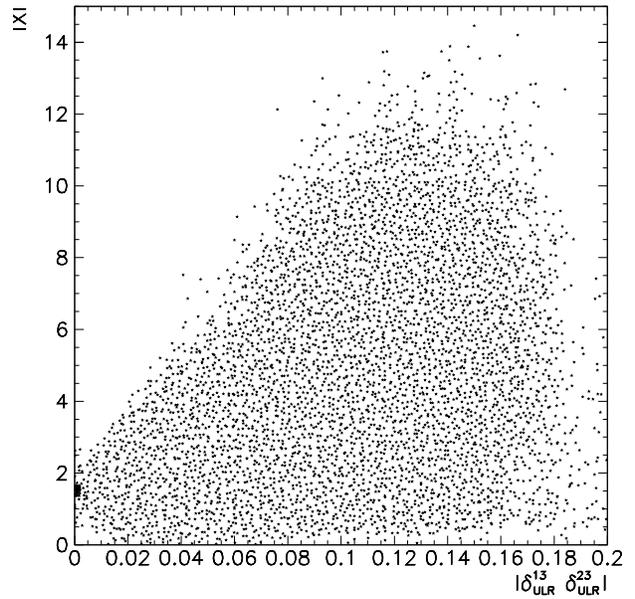}
\caption{SUSY contribution $X$ to ${\cal A}(K \to \pi \nu \bar \nu)$
\cite{Buras:2004qb}
\label{fig:ampKpinunu}       
}
\end{figure}

The decays $K^+ \to \pi^+ \nu \bar \nu$ and $K_L \to \pi^0 \nu \bar
\nu$ are almost unique in that they are essentially free of hadronic
uncertainties. In the SM context, the two modes provide a clean
and independent determination of the unitarity
triangle~\cite{Buchalla:1994tr,Buchalla:1996fp,Isidori:2005xm,Buras:2005gr,Buras:2006gb,Mescia:2007kn,Brod:2008ss}---once
they will have been measured precisely, hopefully, at
CERN P-326/NA62 (formerly NA48/III) and at
JPARC E14.
(The SM branching fractions are ${\cal O}(10^{-11})$
and ${\cal O}(10^{-10})$, respectively.)
They are a precise probe the FCNC vertices of the $Z$ boson.
In view of the particular, $SU(2)$-breaking structure of the
leading $Z s d$ vertex, this implies a specific sensitivity to
certain combinations of $LR$ and $RL$ $\delta$-parameters, even in the
presence of general flavour violation
\cite{Colangelo:1998pm,Nir:1997tf,Buras:1997ij,Buras:1999da,Buras:2004qb,Isidori:2006qy}.
In a perturbative expansion in $\delta$'s,
\begin{equation}
  {\cal A}(K \to \pi \nu \bar \nu)^{\rm SUSY} \propto
    (\delta^\tu_{ut})_{LR} (\delta^\tu_{tc})_{RL} .
\end{equation}
This is an example where the (generically) leading effect arises at second order
in the mass insertions. The double LR mass insertion
enforces a factor $v^2/M^2$ required by the decoupling theorem.
A systematic numerical analysis~\cite{Buras:2004qb}
(see also~\cite{Isidori:2006qy})
shows that this parametric dependence continues to hold beyond the perturbative
expansion, and even in the presence of large contributions from box
diagrams (Fig.~\ref{fig:ampKpinunu}). Moreover the SM hierarchy between the
charged and neutral modes may be reversed, the latter enhanced by an
order of magnitude or more, and the bound following from
isospin symmetry of strong interactions~\cite{Grossman:1997sk} saturated.
Indeed, at present the experimental upper bound provides the best
constraints on the cited double mass insertion.
Complementary probes of the $Z$-penguin amplitude are provided by the
modes $K_L \to  \pi^0 e^+ e^-$ and $K_L \to \pi^0 \mu^+ \mu^-$,
which are still theoretically quite clean~\cite{Isidori:2006qy,Mescia:2006jd}.
Analogous $Z$-penguin effects are possible in $b\to d$
and $b \to s$ transitions, see the next subsubsection and the
brief discussion of EWP effects in Section \ref{sec:bsmhints}.

\subsubsection{Leptonic $B$ decays}
Among $B$-decays, the modes $B^+ \to \ell^+ \nu$ and $B_d, B_s \to \ell^+
\ell^-$ are the theoretically cleanest. The former proceeds
through a $W^+$ tree-level diagram in the SM. Significant corrections
may occur due to charged-Higgs-boson exchange, which is present in the
MSSM~\cite{Isidori:2006pk,Akeroyd:2003zr,Itoh:2004ye} but becomes
relevant only at large values of $\tan\beta$. In this case, the latter
mode can be enhanced by an order of magnitude or more, which provides
a serious constraint on certain large-$\tan\beta$ scenarios. (The
branching fraction scales with the sixth power of $\tan\beta$.)
We would like to emphasize that at any $\tan\beta$,
it will receive more moderate contributions via
the $Z$-penguin, through the operator $Q_{10A}$, which could
still be sizable (compare the discussion in the previous
subsubsection). In the SM one has~\cite{Buras:2003td}
\begin{equation}
   BR(B_s \to \mu^+ \mu^-) = (3.51 \pm 0.50) \times 10^{-9}, 
\end{equation}
where the bulk of the CKM and hadronic uncertainties has been eliminated by
normalizing to $\Delta M_s$. (The error will become even smaller with
improved lattice predictions for $\hat B_{B_s}$.) In spite of this
mode being so rare, LHCb, ATLAS, and CMS expect to collect a combined
few hundred SM events after five years or so of running.

\subsubsection{Inclusive $B \to X_{s} \ell^+ \ell^-$}
Their sensitivity to semileptonic operators like $Q_{9 V}$ and $Q_{10 A}$
makes the rare $b \to s \ell^+ \ell^-$ transitions a complementary and
more complex test of the underlying theory than the radiative ones. The
presence of two charged leptons in the final state allows
to define several observables including dependence on the kinematics.
In the absence of large statistics,
partially integrated spectra such as the dilepton mass spectrum or the
angular distribution can be explored that are amenable to a systematic
theoretical description for a dilepton invariant mass below the charm
resonances.

The decay rate has been considered (in msugra) in \cite{Bertolini:1990if}.
In the (general) minimally flavour-violating MSSM 
\cite{Ali:1994bf,Ali:2002jg,Bobeth:2004jz} the Wilson
coefficients $C_{9 V}$ and $C_{10 A}$
are only slightly affected and corrections to the decay
distributions do not exceed the $30 \%$ level.

At large $\tan \beta$, additional contributions
to $b \to s \mu^+ \mu^-$ arise from the chirality-flipping operators
$(\bar s_L b_R) (\bar \mu_L \mu_R)$ and $(\bar s_L b_R) (\bar
\mu_R \mu_L)$ that are suppressed by powers of the muon mass but
enhanced by $(\tan \beta)^3$. In practice, these contributions are
however bounded from above~\cite{Chankowski:2003wz,Hiller:2003js,Bobeth:2007dw} by the experimental constraints on
$B_s \to \mu^+ \mu^-$ and turn out to be subleading. Merging the
information on $B \to X_s \ell^+ \ell^-$ with the one on $B \to
X_s \gamma$, one can thus infer that the
sign of the $b \to s \gamma$ amplitude should be
SM-like~\cite{Gambino:2004mv} in MFV.
Expressions to the semileptonic Wilson coefficients relevant for
the general MSSM
can be found in \cite{Cho:1996we}. See also \cite{Gabrielli:2002me}.
A simultaneous
use of the $B \to X_s \ell^+ \ell^-$ and $B \to X_s \gamma$
constraints then leads to stringent limits on
$(\delta^\td_{sb})_{LL}$ and $(\delta^\td_{sb})_{LR}$ in the complex
plane~\cite{Ciuchini:2002uv,Silvestrini:2007yf}.

\subsubsection{Exclusive semileptonic and radiative $B$-decays}
Exclusive semileptonic and radiative $B$-decay modes such as
$B \to K^{(*)} l^+ l^-$, $B \to K^* \gamma$, etc.\ are also natural
as a place to look for physics beyond the Standard Model;
they will be (much) more readily accessible than the inclusive
ones in a hadronic environment such as LHCb.
However, the theoretical treatment is somewhat more involved
and they have not received as much attention in a supersymmetric
context as their inclusive counterparts.
See e.g. \cite{Lunghi:2006hc} for a proposal to look for
supersymmetric right-handed currents in $B \to K^* l^+ l^-$.
More work on SUSY effects in these observables would be desirable.

\subsection{Combined constraints} \label{sec:global}
\begin{figure}
\includegraphics[width=0.48\textwidth,angle=0]{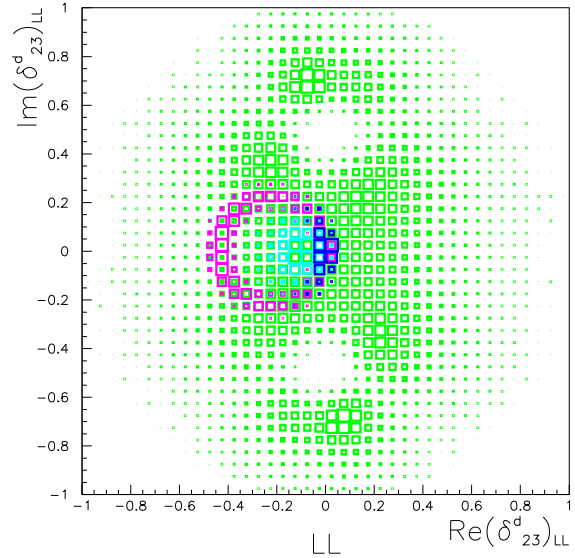}
\caption{Combined constraints on $(\delta^d_{sb})_{LL}$
\cite{Silvestrini:2007yf}. The allowed
regions due to $B \to X_s \gamma$, $B \to X_s \ell^+ \ell^-$, and
$B_s-\bar B_s$ mixing are shown in violet, light blue, and green,
respectively; the combined constraint is in dark blue. A common
sparticle mass of $350$ GeV and $\mu = -350$ GeV is assumed.
\label{fig:combined} }      
\end{figure}

The constraints on the flavour-violating parameters become more
powerful when the interplay of several observables with different
parametric sensitivities is considered. (This was considered in
a number of the works referred to above.)
For $b \to s$ transitions, combined constraints on
the $\delta_{sb}$ parameters were obtained
and graphically displayed in ref. \cite{Silvestrini:2007yf}.
Fig. \ref{fig:combined} \cite{Silvestrini:2007yf} shows 
how the measurements of from $B \to X_s \gamma$, $B \to X_s \ell^+
\ell^-$, and $\Delta M_s$ coact to constrain the parameter 
$(\delta^d_{sb})_{LL}$, leaving a much smaller allowed region than
each individual observable.

\subsection{QCD penguins}

QCD-penguin graphs contribute only to $\Delta F=1$
transitions via the operators
\begin{eqnarray}  \label{eq:chromomag}
  Q_{8g} &=& - \frac{g_s\, m_b}{4 \pi^2}
        \bar s_{L}\, \sigma^{\mu \nu} b_{R} G_{\mu\nu}, \\
  Q_{3,4} &=& \sum_q (\bar s_{L} \gamma^\mu b_{L})\, (\bar q_L \gamma_\mu q_L), \\
  Q_{5,6} &=& \sum_q (\bar s_{L} \gamma^\mu b_{L})\, (\bar q_R \gamma_\mu q_R),
\end{eqnarray}
together with operators $Q_i'$ which arise from the $Q_i$ by changing
the chiralities of all quarks.
Here we have taken the case of $b \to s$ transitions as an example,
with obvious replacements for $b \to d$ and $d \to s$
transitions. We have also suppressed colour and part of the Dirac
structure. ($G$ is the gluon field strength.) Note that the QCD
penguins $Q^{(')}_{3 \dots 6}$ involve $b$ and $s$ quarks of like
chiralities, while the chromomagnetic penguins $Q^{(')}_{8 g}$
involve a chirality flip. As with the electroweak
penguins, this engenders specific patterns of
sensitivity of their coefficients to the parameters $\delta$, e.g.\ 
$(\delta^\td_{sb})_{LR}$ in the case of  $C_{8g}$.

\subsubsection{Charmless hadronic decays}
Two-body exclusive nonleptonic decays $B \to M_1 M_2$ are sensitive to
all of the operators $Q^{(')}_{1\dots10,7\gamma, 8g}$, while offering a
large number ${\cal O}(100)$ of observables, including many CP-violating
ones. They (and exclusive modes in general) will increase
in importance due to their accessibility at hadron machines such as the LHC.
On the theoretical side, the factor limiting the precision are the
hadronic matrix elements $\langle M_1 M_2 | Q_i | B \rangle$, which
involve nonperturbative QCD in a way that is presently not
surmountable in lattice QCD. 
Systematic methods are, however, available, based on expansions
about the limit of $SU(3)$ flavour symmetry or about the
heavy-$b$-quark limit. $m_s/\Lambda$ and $\Lambda/m_b$, respectively,
are the expansion parameters. In fact, the (QCD)
factorization formulae
\cite{Beneke:1999br,Beneke:2000ry,Beneke:2001ev}
for the hadronic matrix elements that follow
from the heavy-quark limit respect the $SU(3)$ flavour symmetry up
to well-defined (so-called ``factorizable'') corrections at the
leading power and perturbative order, and perturbative QCD corrections
do not alter this picture very much. Higher orders in $\Lambda/m_b$
are generally not under control, with certain (important) exceptions.
We cannot go into more detail here but refer to the brief discussion
of $b \to s$ penguin modes in Section \ref{sec:bsmhints}.

\subsection{Lepton flavour violation}
It is beyond the scope of this article
to cover lepton flavour violation in detail.  In the SM,
even when introducing the minimal dimension-five operator to allow for
neutrino mass\-es and mixings, lepton-flavour-violating processes such
as $\tau \to \mu \gamma$ are rendered extremely rare by the
tiny neutrino mass splittings and the corresponding near-perfect
GIM cancellation. The situation is very different in the MSSM because
of the presence of lepton flavour violation at the renormalizable
level, in the sneutrino and charged slepton mass matrices.
Nonobservation of SUSY effects in $\ell_i \to \ell_j \gamma$ leads
to stringent bounds on sleptonic mass insertions
\cite{Gabbiani:1996hi}, ruling out generic flavour structures for
them. Predictions for MSSM (charged) lepton flavour violation are
often considered in seesaw scenarios. See \cite{Hisano:2008ry} for a
recent review, and Section \ref{sec:susygut}

\subsection{Implications of low-energy observables for 
collider-physics measurements}
The flavour-violating parameters that govern low-energy observables
also affect production cross sections for sparticles, and their
decays; a detailed investigation would go beyond the scope of the
present review. For recent works in that direction, see
e.g. \cite{Bozzi:2007me,Dittmaier:2007uw,Feng:2007ke,Fuks:2008ab,Hiller:2008wp,Buchmueller:2008qe}, and \cite{Buchalla:2008jp} (which
also deals with many of the flavour observables discussed above)
and references therein.

\section{Probing the GUT scale} \label{sec:susygut}
\begin{figure}
\begin{center}
\parbox{74mm}{\vskip3mm
\psfragscanon
\psfrag{m6}{$10^{-6}$}
\psfrag{mexp}{$6 \cdot 10^{-7}$}
\psfrag{m6p5}{$10^{-6.5}$}
\psfrag{m7}{$10^{-7}$}
\psfrag{m7p5}{$10^{-7.5}$}
\psfrag{m8}{$10^{-8}$}
\psfrag{m9}{$10^{-9}$}
\psfrag{mgl}{$m_{\tilde g_3}$}
\psfrag{msq}{\Large $m_{\tilde q}$}
\psfrag{adbymsq}[Bl][Bl][1][0]{\Large $a_{d}/m_{\tilde q}$}
\resizebox{74mm}{!}{\includegraphics{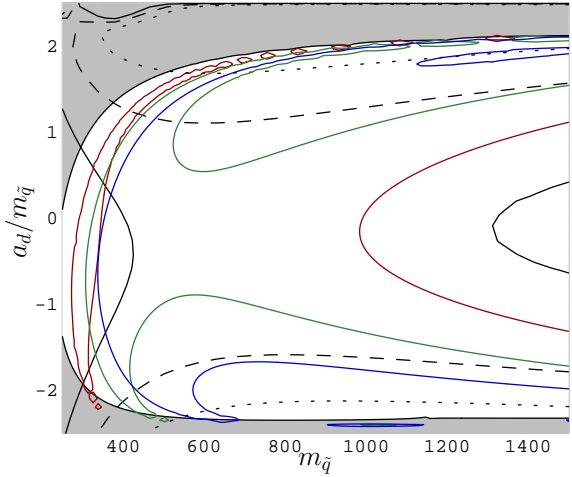}}
}
\end{center}
\caption{Contours of constant SUSY contribution to $\Delta M_{B_s}$,
  in units of the SM value, and of constant
  $BR(\tau\to\mu\gamma)$, in a slice of parameter space \cite{Jager:2005ii}.
  Here $a_d$ has a simple relation to $a_0$, $m_{\tilde g} = 250$ GeV,
  $\tan\beta=3$. The solid black, dashed, and dotted contours
  correspond to $|\Delta M_{B_s}^{\rm (NP)}|/\Delta M_{B_s}^{\rm (SM)} = 0.5$,
  $2$, $5$, respectively. The red, green, and blue contours correspond
  to the experimental upper bound on $BR(\tau \to \mu \gamma)$ for
  $\mu=-300$, $\mu=-450$, $\mu=-600$ GeV, respectively. ($\Delta
  M_{B_s}$ is independent of $\mu$ in the approximation used.)}
\label{fig:dmbrtmg}       
\end{figure}

Concrete assumptions about the SUSY-breaking mechanism (gravity
mediation, gauge mediation, etc.) and possible UV completion (such as a SUSY
grand-unified theory, minimal flavour violation, etc.) may imply
patterns in ${\cal L}_{\rm soft}$ relating different $\delta$ parameters
that can be tested against the general constraints applying to them,
or can be further used in making specific predictions for low-energy
observables and their correlations. We do not discuss specific
SUSY models in this article, referring instead to other
contributions to this volume~\cite{Raby:2008gh,Nilles:2008gq} for
the status of SUSY model building, but consider only a very minimal
GUT setup for illustration.
One of the most intriguing aspects of SUSY GUTs,
which also demonstrates the power of flavour
observables to probe even superhigh scales, is the possibility of
relations between hadronic and leptonic flavour violation
\cite{Barbieri:1994pv,Barbieri:1995tw,Barbieri:1995rs,Hisano:1998fj,Baek:2000sj,Moroi:2000mr,Moroi:2000tk,Akama:2001em,Chang:2002mq,Masiero:2002jn,Hisano:2003bd,Goto:2003iu,Ciuchini:2003rg,Jager:2003xv,Jager:2004hi,Jager:2005ii,Grinstein:2006cg,Ciuchini:2007ha,Albrecht:2007ii,Parry:2007fe,Dutta:2008xg,Hisano:2008df}.
The effect we are considering here is the
following~\cite{Hall:1985dx,Barbieri:1995tw,Barbieri:1995rs}. Assume
that SUSY breaking is effected at a scale beyond the GUT scale, for
instance at the Planck scale, and that it is
flavour-blind, at least approximately, such that one has a
universal scalar mass parameter $m_0^2$ and a universal trilinear
scalar coupling parameter $a_0$. For definiteness, assume
simple $SO(10)$ unification such
that there is only one sfermion multiplet for each
generation. Radiative corrections due to the unified gauge coupling
will correct the masses of the three $16$'s of sfermions in the same
way, while the large top Yukawa coupling will selectively suppress the masses of
one multiplet,
\begin{equation}   \label{eq:so10shift}
    m^2_{16_{1,2}} = m_0^2 + \epsilon, \qquad
    m^2_{16_3} = m_0^2 + \epsilon - \Delta ,
\end{equation}
where $\epsilon \propto g^2$ and $\Delta \propto y_t^2 m_0^2,\, y_t^2 a_0^2$. 
Eq.~\eq{eq:so10shift} holds in a basis where the up-type Yukawa matrix
is diagonal.
(Further contributions will be present if there are additional
large Yukawa couplings, such as for large $\tan\beta$. The expressions
then become more complicated.) In this fashion, the large top Yukawa
coupling affects also the right-handed down-type squark masses and the
slepton masses, which is very different from the situation in the MSSM
(or below the GUT scale). Now the relevant sfermion basis for low-energy
physics (the super-CKM basis) is the one where the down-type and
leptonic Yukawas are diagonal. In such a basis the members of the
$16_3$ selected by $y_t$ will consist of mixtures of the superpartners
of fermions of different flavours, and FCNC SUSY vertices appear.
Moreover, these vertices will be correlated between the hadronic and
leptonic sectors.
This kind of mechanism received renewed attraction after the
observation of large leptonic mixing in neutrino
oscillations~\cite{Chang:2002mq}. Assuming
a minimal-$SU(5)$-type embedding of the MSSM into $SO(10)$, and under
certain assumptions about the generation of seesaw neutrino masses,
one has
\begin{equation}
   Y_D = U_{\rm PMNS}^T\, {\rm diag}(y_d, y_s, y_b)\, V_{\rm CKM}^\dagger.
\end{equation} 
This expression should be most robust in the $(2,3)$ sub-block. The
atmospheric neutrino mixing angle then appears in the gluino-squark
couplings
\begin{equation}
  {\cal L}_{\rm soft} \supset U_{\rm PMNS}^{ij} \, {\tilde d}^*_{Rj} \,
  \tilde g T^A \, d_{Ri} ,
\end{equation}
yielding potentially spectacular effects in observables like
$\Delta M_{B_s}$
\cite{Harnik:2002vs,Jager:2003xv,Jager:2004hi,Jager:2005ii}.
The model is parameterized by four parameters $m_0$, $a_0$, $m_{\tilde
  g}$, and $\mu$, and a value of
$\tan\beta$ around $2-3$ to maintain perturbativity. Fig.~\ref{fig:dmbrtmg}
compares $BR(\tau \to \mu \gamma)$ with the correction to $\Delta
M_{B_s}$. It is evident that the measurement of $\Delta M_{B_s}$
provides a nontrivial and quantifiable constraint on the
lepton-flavour-violating mode, providing (one of many) illustration(s)
of the possibility to probe very fundamental scales with
flavour-violating observables, beyond what is possible with knowing
the particle spectrum alone.

\section{Hints of departures from the standard model}
\label{sec:bsmhints}
\subsection{CP violation in $B_s - \bar B_s$ mixing?} \label{sec:bsmphis}
We mentioned in Section \ref{sec:bbmix} that the Standard Model
predicts, in a theoretically very clean manner,
negligible mixing-induced CP violation in $B_s$ decays,
and also the Tevatron measurements of time-dependent CP violation
in $B \to J/\psi \phi$, which after 2008 summer conferences
shows an (increased) $2.2$ $\sigma$ discrepancy with the SM
expectation. In fact, as mentioned
in Section \ref{sec:bbmix}, further information on $\phi_s$ can be
inferred from the ``semileptonic CP asymmetry'' or similar
asymmetries in other flavour-specific $B_s$ decays. Combined fits
have been performed by the UTfit and CKMfitter collaborations.
This raises the
discrepancies with the SM in $\phi_s$ to $2.5$ and $2.4$ $\sigma$
\cite{Bona:2008jn,PieriniIchep08,DeschampsIchep08},
respectively. Recall that the expected precision after 5 years of LHCb
is about 1 degree, hence this has the potential to turn into a very
significant falsification of the SM. It is not difficult to generate
such a signal in the MSSM, as $b \to s$ FC parameters are not very
strongly constrained \cite{Parry:2007fe,Dutta:2008xg,Hisano:2008df}.
These papers also relate the effects in $B_s-\bar B_s$ mixing to
the branching fraction for $\tau\to\mu \gamma$ in the context of
SUSY GUTs of the type discussed in Section \ref{sec:susygut}.

\subsection{Time-dependent CP asymmetries (``$\sin 2 \beta_{\rm  eff}$'') }
\begin{figure}
\begin{center}
\includegraphics[width=0.43\textwidth,angle=0]{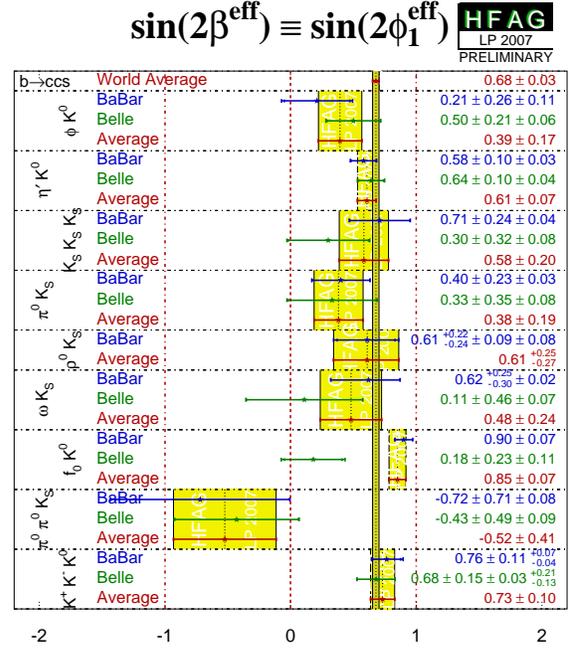}
\caption{Mixing-induced CP violation in charmless $b\to s$ penguin
  transitions \cite{Barberio:2007cr}.}
\label{fig:sin2beff}       
\end{center}
\end{figure}
The most long-standing pattern of observed deviations in flavour physics
concerns the closely related class of
time-dependent CP asymmetries in $b\to s$ penguin decays of
$B_d$ mesons into CP eigenstates,
\begin{eqnarray}
\lefteqn{  \frac{BR(\bar B^0(t)\to f) - BR(B^0(t) \to f)}
       {BR(\bar B^0(t)\to f) + BR(B^0(t) \to f)} } \nonumber \\
  & & \equiv S_f \sin(\Delta m_B t) - C_f \cos(\Delta m_B t) .
\end{eqnarray}
Within the SM one expects $ \eta_f S_f \approx \phi_d = 2 \beta $,
based on the dominance of the QCD penguin amplitude, which has
vanishing weak phase. ($\eta_f$ denotes the CP quantum number of the
final state.) Neither equality holds in the MSSM (beyond minimal
flavour violation). Fig.~\ref{fig:sin2beff} shows
the data for a number of modes. It is conspicuous that in general, the
$\eta_f S_f \equiv \sin(2\beta^{\rm eff})(f)$ lies below the value of
$\sin 2\beta$.This has received much interest in recent years.
Unfortunately, these modes are theoretically less clean for
new-physics searches, as they require the calculation of at least
partial information  about hadronic decay amplitudes
(at the minimum, restricting a strong phase to a half-plane).
Moreover, the significance of the ``signal''
has gone down over the past few years. Perhaps we are observing
merely a statistical fluctuation. Nevertheless, any reasonably SUSY
model that can accomodate CP violation in $B_s -\bar B_s$ mixing
($\Delta B=\Delta S=2$) should give rise to {\em some} contribution
to ($\Delta B=\Delta S=1$) $b\to s$ penguin transitions, since the
fundamental FC vertices in the MSSM all violate flavour numbers by one.

In fact, in the SM a QCD factorization calculation~\cite{Beneke:2005pu}
of the corrections
for the two-body final states $\phi K^0$, $\eta' K^0$, $\pi^0 K_S$,
$\rho^0 K_S$, $\omega K_S$ due to neglected smaller amplitudes
shows a (small) positive shift in all cases except $\rho^0 K_S$.
Supersymmetric contributions to $b \to s$ hadronic penguin transitions
have received considerable attention in the past
\cite{Borzumati:1999qt,Ciuchini:2002uv,Bertolini:1987pk,Grossman:1996ke,Ciuchini:1997zp,Barbieri:1997kq,Kagan:1997sg,Lunghi:2001af,Causse:2002mu,Khalil:2002fm,Khalil:2003bi,Agashe:2003rj,Kane:2002sp,Kane:2003zi,Harnik:2002vs,Chakraverty:2003uv,Dariescu:2003tx,Ball:2003se,Khalil:2004yb,Gabrielli:2004yi},
under various assumptions about where flavour is violated.
Often the most important operators are the magnetic-penguin
operators $Q_{8g}$ or $Q'_{8g}$, which are sensitive to $LR$
$\delta$-parameters. ($Q'_{8g}$ will interfere with the SM contributions in
exclusive processes, unlike in inclusive $B \to X_s \gamma$ decay,
which makes it less constrained but possibly relevant.)
Indeed, an interesting possibility is to attribute the pattern of the
deviations to constructive and destructive interference between
operators of different quark chiralities, depending on the parities of
the final-state particles~\cite{Khalil:2003bi}.
Another possibility to change the $b \to s$ penguin pattern from its
SM form is a modified electroweak penguin, see below.

\subsection{Pattern of CP asymmetries in $B \to \pi K$}
$B \to \pi K$ decays are those charmless hadronic decays that have
been studied most with regard to possible new-physics effects. 
(This set of observables overlaps with those of the previous
subsection in the mixing-induced CP violation parameter
$S_{\pi^0K_S}$). 
This interest relates to the fact that they are penguin-dominated, receiving
important contributions from QCD penguins, as well as
electroweak-penguin contributions that one can try to disentangle
using an isospin analysis. Also here, some knowledge of had\-ronic
amplitude ratios is needed. Three ``puzzles'' have been identified.
\begin{figure}
\begin{center}
\includegraphics[width=0.44\textwidth,angle=0]{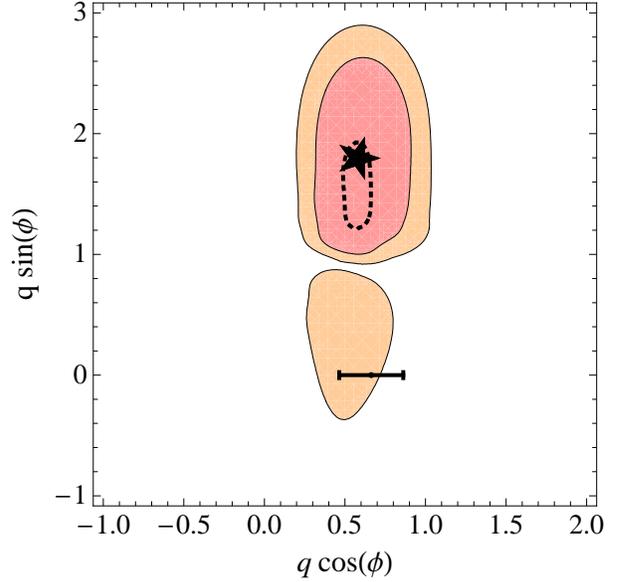}
\caption{Electroweak-penguin-to-tree ratio fitted from data
\cite{Fleischer:2008wb}. Shown are $1\sigma$ and $90$ \% CL regions.
The star denotes the best fit, the bar the SM expected range,
which lies along the real line. For details, see \cite{Fleischer:2008wb}.}
\label{fig:ewppar}       
\end{center}
\end{figure}
\subsubsection{Ratios of isospin-related branching fractions}
While the deviation of the pattern of branching fractions from SM expectations
observed in \cite{Buras:2003dj,Yoshikawa:2003hb,Gronau:2003kj,Beneke:2003zv}
has gone away as the experimental data (and its treatment) have
evolved, the puzzle invoked a
number of theoretical calculations on the MSSM impact, see
in particular \cite{Khalil:2004yb}, as well as the references in the
previous subsection, which may be relevant to the next two issues.
\subsubsection{Patterns in direct CP asymmetries}
Theoretical arguments (such as
calculations based on the heavy-quark limit) suggest that
\begin{eqnarray}
   &&  C_{\pi^- K^+} \equiv - A_{CP}(B_d^0 \to \pi^- K^+) \nonumber \\
   && \qquad     \approx C_{\pi^0 K^+} \equiv - A_{CP}(B^+ \to \pi^0 K^+)
\end{eqnarray}
in the Standard Model. The HFAG-averaged data
due to Babar, Belle, CDF and CLEO is \cite{Barberio:2007cr}
$$
   C_{\pi^- K^+} = 0.097 \pm 0.0012,
   \qquad C_{K^+ \pi^0} = -0.050 \pm 0.025 ,
$$
a $>5$ $\sigma$ difference \cite{:2008zza,Peskin:2008zz}.
The theory expectation (in the SM) is hard to quantify precisely,
mainly due to the uncertain colour-suppressed tree contribution,
but $0.15$ appears very large. One mode depends on electroweak penguins
but not the other, which makes the difference sensitive to possible,
CP-violating, new-physics effects.
 
\subsubsection{Non-standard electroweak penguin and $S_{\pi^0 K_S}$}
An alternate route is to use SM isospin relations and use the data
on the rates to predict the parameter $S_{\pi^0 K_S}$. This allows
using theory input in a more limited fasion. Fig. \ref{fig:ewppar}
shows the result of a $\chi^2$ fit of the relevant electroweak penguin
parameter to data \cite{Fleischer:2008wb}.
Also here, the significance is not very strong. Again, however, the
data are certainly consistent with extra SUSY contributions, and we
expect on general grounds some correlation with any new-physics
effects in $B_s-\bar B_s$ mixing.

\section{Conclusion}
The MSSM contains a large, ${\cal O}(100)$, number of new
flavour and CP parameters. We have emphasized their tight connection
with the mechanism of supersymmetry breaking. The MSSM flavour structures
have to be quite non-generic both in the quark and in the
lepton sector. However,
there is room for deviations from minimal flavour violation. In fact,
several possible hints of NP are present in the data.
In the next few years, only LHC can tell us whether they are real
or fluctuations.

As this is being written, CERN has just shot its first test bunches into
the LHC ring. We may hope that soon the era of putting bounds
and constraints will give way to a phase of actual measurements of MSSM
parameters.

If SUSY is found, the interplay between direct and indirect observables
is likely to be useful, as it was in the construction of the SM.
A super-$B$-factory will be of great help here.
Moreover, the peculiar gentleness of SUSY quantum corrections
may mean that in SUSY, the GUT or Planck scales can actually
be ``close'' as far as indirect observables are considered (as
exemplified in certain SUSY GUTs), even if being at great
distance from the point of view of direct detection.

\section*{Acknowledgement}
I am grateful to G. Honecker for discussions and to J.~Zupan,
P. Slavich, and C. Smith for conversations and helpful comments on
the manuscript. The author is supported in part by the RTN European Program
MRTN-CT-2004-503369.
\appendix
\section{Loop functions} \label{app:loopfun}
\begin{equation}
   S(x) = \frac{4 x - 11 x^2 + x^3}{4(1-x)^2}
         - \frac{3\, x^3 \ln x}{2 (1-x)^3}
\end{equation}
\begin{eqnarray}
   f_6(x) &=& \frac{6\,(1+3x) \ln x + x^3 - 9 x^2 - 9 x + 17}{6(x-1)^5} \\
   \tilde f_6(x) &=& \frac{6\,x(1+x) \ln x - x^3 - 9 x^2 + 9 x + 1}{3 (x-1)^5}
\end{eqnarray}

%
%

\end{document}